\documentclass{emulateapj}
\newcommand{\Sgra}{Sgr A$^{*}$}
\newcommand\Msun{\ensuremath{\mathrm{M_\odot}}}
\newcommand\Lsun{\ensuremath{\mathrm{L_\odot}}}
\usepackage{epstopdf}
\usepackage{amsmath}
\usepackage{bbding}
\usepackage{color}
\begin{document}
\title{The Post-Periapse Evolution of Galactic Center Source G1: The second case of a resolved tidal interaction with a supermassive black hole}
\author{\sc G. Witzel\altaffilmark{1}, B. N. Sitarski\altaffilmark{1}, A. M. Ghez\altaffilmark{1}, M. R. Morris\altaffilmark{1}, A. Hees\altaffilmark{1}, T. Do\altaffilmark{1}, J. R. Lu\altaffilmark{2}, S. Naoz\altaffilmark{1}, A. Boehle\altaffilmark{1}, G. Martinez\altaffilmark{1}, S. Chappell\altaffilmark{1}, R. Sch\"{o}del\altaffilmark{3}, L. Meyer\altaffilmark{1}, S. Yelda\altaffilmark{1}, E. E. Becklin\altaffilmark{1,4}, K. Matthews\altaffilmark{5}}

\email{bsitarski@astro.ucla.edu}

\altaffiltext{1}{Department of Physics and Astronomy, University of California, Los Angeles, 430 Portola Plaza, Los Angeles, CA 90095-1547, USA}
\altaffiltext{2}{Institute for Astronomy, University of Hawaii, 2680 Woodlawn Drive, Honolulu, HI 96822, USA}
\altaffiltext{3}{Instituto de Astrofisica de Andalucia (CSIC, Glorieta de la Astronomia S/N, 18008 Granada, Spain}
\altaffiltext{4}{NASA-Ames Research Center, MS 232-12, Moffett Field, CA 94035, USA}
\altaffiltext{5}{Division of Physics, Mathematics, and Astronomy, California Institute of Technology, Pasadena, CA 91125, USA}

\begin{abstract}

We present new Adaptive Optics (AO) imaging and spectroscopic measurements of Galactic Center source G1 from W. M. Keck Observatory. Our goal is to understand its nature and relationship to G2, which is the first example of a spatially-resolved object interacting with the supermassive black hole (SMBH). Both objects have been monitored with AO for the past decade (2003 - 2014) and are comparatively close to the black hole ($a_{\rm{min}} \sim$200-300 AU) on very eccentric orbits ($e_{\rm{G1}}\sim$0.99; $e_{\rm{G2}}\sim$0.96). While G2 has been tracked before and during periapse passage ($T_{0} \sim$ 2014.2), G1 has been followed since soon after emerging from periapse ($T_{0} \sim$ 2001.3). Our observations of G1 double the previously reported observational time baseline, which improves its orbital parameter determinations. G1's orbital trajectory appears to be in the same plane as that of G2, but with a significantly different argument of periapse ($\Delta\omega$ = 21$\pm$4 degrees). This suggests that G1 is an independent object and not part of a gas stream containing G2 as has been proposed. Furthermore, we show for the first time that: (1) G1 is extended in the epochs closest to periapse along the direction of orbital motion and (2) G1 becomes significantly smaller over time, (450 AU in 2004 to less than 170 AU in 2009). Based on these observations, G1 appears to be the second example of an object tidally interacting with a SMBH. G1's existence 14 years after periapse, along with its compactness in epochs further from the time of periapse, suggest that this source is stellar in nature.

\end{abstract}

\section{Introduction}

As the capabilities of high-resolution imaging facilities have advanced, the center of our Galaxy has become a unique laboratory for studying the nearest supermassive black hole (SMBH; \citealt{ghez1998, ghez2008, gillessen2009}) and has revealed many unexpected results. This includes the presence of young stars where none were expected (e.g., \citealt{levin2003, genzel2003,paumard2009, bartko2009, lu2009, yelda2014}), a lack of old stars where many were predicted (e.g., \citealt{buchholz2009, do2009b}), and very faint, but highly variable, infrared emission believed to be associated with the black hole's accretion flow. 

The most recent Galactic Center discovery from high-resolution infrared observations that has attracted considerable attention is the very red, infrared source G2, which recently went through closest approach where its tidal interaction should have been maximal ($T_{0, \rm{G2}}$ = 2014.21 $\pm$ 0.13; \citealt{meyer2014}). It was originally hypothesized to be a 3 Earth-mass gas cloud, and as it went through closest approach to the supermassive black hole, Sgr A$^{*}$, it was projected to tidally disrupt, shock and to possibly contribute to an enhanced accretion episode onto the black hole \citep{gillessen2012, burkert2012, schartmann2012, pfuhl2015}. Observations of G2 after periapse passage have challenged the gas cloud hypothesis. First, it survived as a compact source in the continuum imaging measurements at 3.8 $\mu$m \citep{witzel2014} and possibly as a compact source in the gas (Br-$\gamma$ spectroscopic measurements; \citealt{valencia2015}). This has favored the alternative hypothesis that there is a central stellar source embedded in G2. There are several variants of the stellar hypothesis, including: a disrupted protoplanetary disk \citep{murrayclay2012}; a disrupted disk around an old star \citep{miralda2012}; a mass-loss envelope from a young T Tauri star \citep{scoville2013}; a Wolf-Rayet star \citep{eckart2013}; spherically symmetric winds from an embedded object \citep{ballone2013}; a binary merger product \citep{phifer2013, witzel2014, prodan2015}; and an embedded pre-main sequence star \citep{valencia2015}.

More recently, another object -- G1 -- has been recognized to bear a close relationship to G2. G1 was originally found to be another very red, extended infrared source that was interpreted as a spatially-resolved, stationary hot dust feature that is locally heated by nearby stars surrounding Sgr A$^{*}$ \citep{clenet2004, clenet2005, ghez2005a}. In addition, \citet{pfuhl2015} noted that G1 has observational properties similar to those of G2, including Br-$\gamma$ emission as well as a very red color. Also, G1 passed through periapse $\sim$13 years ago \citep{pfuhl2015, sitarski2014}, and therefore high-resolution observations only exist \textit{post}-periapse passage while we have observations of G2 prior to, through, and post-periapse passage. The observations in \citet{pfuhl2015} also suggest that G1 moves on a Keplerian orbit with orbital characteristics similar to G2 \citep{sitarski2014}. These similar orbits and observational characteristics led \citet{pfuhl2015} to hypothesize that G2 and G1 are part of a gas streamer on the same trajectory.

In this paper, we explore the evolution of G1's observed properties and orbital motion over the last decade, the longest time baseline reported thus far for this object. We investigate the evolution of G1 with time and position from Sgr A$^{*}$ to characterize its tidal interactions. With our longer time baseline, we test the theory that G1 and G2 are part of the same gas streamer.

This paper is organized as follows: Section~\ref{datasets} describes our data sets and data reduction techniques; Section~\ref{analysis} details our astrometric and photometric calibration, and our orbital fitting procedure; Section~\ref{results} presents our results; and Section~\ref{discussion} discusses our findings in the context of G2 and evidence that these are self-gravitating objects. One scenario that we consider is the binary merger hypothesis. Section~\ref{concl} summarizes our conclusions.

\section{Data Sets}\label{datasets}
Near-infrared, high-angular-resolution images and spectra of the Galactic Center region containing G1 have been obtained as part of the long-term program at the W. M. Keck Observatory (WMKO), carried out by our group, to study the Galactic Center black hole and its environs. In this paper, the primary data sets are WMKO images that have been acquired through the $\rm{L}^{\prime}$ ($\lambda_{0}$ = 3.8 $\mu$m) broadband filter over a thirteen-year period with NIRC2, the facility near-infrared camera (PI: K. Matthews) fed by the Keck II laser guide star adaptive optics system (LGSAO; \citealt{wizinowich2006, vandam2006}). Ten of the twelve epochs have been previously reported by us and are part of our group's archive of fully calibrated data sets \citep{ghez2004, ghez2005a, hornstein2007,phifer2013,witzel2014}. Two additional epochs of observation, 2013 August and 2016 May, are reported here for the first time. The pixel scale for these data sets is 9.950 mas/pixel \citep{yelda2010}, which corresponds to an oversampling factor of $\sim$9 for typical point spread function. Tab.~\ref{obssum} summarizes all the $\rm{L}^{\prime}$ imaging data sets for this study.

The new $\rm{L}^{\prime}$ data sets were observed and calibrated using the same techniques described in the papers reporting our other $\rm{L}^{\prime}$ data sets \citep{stolte2010,phifer2013,witzel2014}. This followed standard techniques with the exception of the treatment of the sky exposures, which were taken for each field rotator mirror position within the same range as the science data in steps of $\sim$2 degrees. For each $\rm{L}^{\prime}$ science exposure in epochs after 2004, the corresponding sky exposure was subtracted in order to accurately subtract the thermal emission from dust on the mirror optics (e.g., \citealt{stolte2010}). Once the data were fully calibrated, selected frames were combined into an average map (main map). The individual frames were selected based on the image quality as measured by the full-width at half-maximum of the PSF (FWHM$\le$1.25$\times$FWHM$_{\textrm{min}}$, where FWHM$_{\textrm{min}}$ is the minimum measured FWHM of all the data) and were weighted by the Strehl ratio of each image. We additionally created three independent images (sub-maps) from three interleaved subsets of frames to determine astrometric and photometric uncertainties for the images. 

For this study, we also draw upon two other types of imaging data sets. The first are two Ms ($\lambda_{0}$ = 4.67$\mu$m) observations obtained on 2005 July 16 (previously published by \citealt{hornstein2007}) and another obtained on 2016 May 21. These were added to enhance our photometric characterization of G1. Second, we used all of our group's $\rm{K}^{\prime}$ data sets, which cover the central 10" $\times$ 10" of our Galaxy, and that have been obtained to track the orbital motions of stars at the Galactic Center \citep{ghez1998, ghez2000, ghez2003, ghez2005b, ghez2008, lu2009, yelda2010, meyer2012, yelda2014, boehle2014}.  In addition to the previously published $\rm{K}^{\prime}$ data sets, two new data sets, obtained on 2013 July 20 and 2016 May 21, are included in this work. The first data set was taken in an identical way to all previous $\rm{K}^{\prime}$ astrometric maps (e.g., \citealt{yelda2014}), consists of 193 frames of data, and its final combined image has a point spread function with a FWHM of 58.5 mas and a Strehl ratio of 0.36. The second was taken similarly to our $\rm{L}^{\prime}$ observations, in which we stared at a fixed position in the central field to minimize overheads. This map consists of 21 frames of data, and its final combined image has a PSF with a FWHM of 68 mas and a Strehl ratio of 0.26. The $\rm{K}^{\prime}$ data are used for both the photometric and astrometric characterization of G1.

Additionally, we utilize a spectroscopic data set obtained at W. M. Keck Observatory with OSIRIS \citep{larkin2006} This data set, which was originally published in \citet{ghez2008}, consists of 28 frames taken 2006 June 18 and 30 and 2006 July 01 through the narrow-band Kn3 filter ($\lambda_{0}$ = 2.166 $\mu$m) with the 35 mas pixel scale. These observations have a spatial resolution at Br-$\gamma$ of 67 mas and a spectral resolution of $\sim$3600. These OSIRIS data constitute some of our deepest Kn3 observations prior to 2012\footnote{OSIRIS was moved from Keck 2 to Keck 1 in January 2012 and the grating was upgraded in January 2013; there have been small reduction artifacts that affect the detection of faint emission-line objects in crowded fields.}, with a total integration time of $\sim$7 hours, and while at that time G1 was near Sgr A$^{*}$, it was sufficiently separated ($r$ = 0.114 $\pm$ 0.009 arcseconds) for the position of Sgr A$^{*}$ and G1 to be disentangled.

\begin{turnpage}
\begin{deluxetable*}{llllllllllllllllllllllllllllll}
\tabletypesize{\scriptsize}
\tablecaption{Summary of Keck/NIRC2 $\rm{L}^{\prime}$ ($\lambda_{0}=3.8 \mu$\textrm{m}) Data \label{obssum}}
\tablewidth{0pt}
\tablehead{\colhead{UT Date} & \colhead{Decimal} & \colhead{t$_{\textrm{int}} \times$ coadds} & \colhead{Frames} & \colhead{Frames} & \colhead{Array Size} & \colhead{Dithered} & \colhead{PSF FWHM} & \colhead{Strehl} & \colhead{$\rm{L}^{\prime}_{\textrm{lim}}$\tablenotemark{a}} & \colhead{$\delta_{x}$}  & \colhead{Original}\\
& \colhead{Date} & & \colhead{Taken} & \colhead{Used} & &  \colhead{FOV} & \colhead{(mas)} & & \colhead{(mag)} & \colhead{(mas)} & \colhead{Publication\tablenotemark{b}}}
\startdata
2002 May 31 & 2002.413 & 0.50 $\times$ 40 & 53 & 25         & 10" $\times$ 10 "                   & 11\farcs6 $\times$ 10\farcs5 & 100 & 0.26 & 12.0 & 2.2 & 0\\ 
2003 Jun 10 & 2003.440    & 0.50 $\times$ 40  & 12 &  6       & 10" $\times$ 10"                    & 12\farcs8 $\times$ 10\farcs2 &  85   & 0.41 & 13.3 & 1.4 & 0\\
2004 Jul 26 & 2004.567      & 0.25 $\times$ 120 & 11 & 11     & 10" $\times$ 10"                    & 10\farcs7 $\times$ 9\farcs8    & 80   & 0.42 & 14.4 & 0.4 & 1\\
2005 Jul 30 \& 31 & 2005.580      & 0.50 $\times$ 60  & 62 & 56     & 10" $\times$ 10"                    & 10\farcs3 $\times$ 9\farcs5    & 81   & 0.36 & 14.4 & 0.3 & 1\\
2006 May 21 & 2006.385    & 0.50 $\times$ 60  & 19 & 19     & 10" $\times$ 10"                    & 11\farcs4 $\times$ 11\farcs3  & 82   & 0.38 & 14.4 & 0.4 & 2\\
2009 Jul 22 & 2009.556      & 0.50 $\times$ 60  & 4 & 4       & 10" $\times$ 10"                    & 13\farcs1 $\times$ 12\farcs3  & 85   & 0.38 & 13.2 & 0.1 & 3\\
2012 Jul 20-23 & 2012.551 & 0.50 $\times$ 30  & 1316 & 1231 & 2\farcs64 $\times$ 2\farcs64\tablenotemark{c} & 2\farcs6 $\times$ 2\farcs8     & 92    & 0.51 & 15.3 & 0.2 & 3\\
2013 Aug 13 & 2013.616     & 0.50 $\times$ 60  & 249 & 245   & 2\farcs64 $\times$ 2\farcs64\tablenotemark{c} & 2\farcs7 $\times$ 2\farcs6     & 90    & 0.54 & 15.2 & 0.1 & 4\\
2014 Mar 20 & 2014.216     & 0.50 $\times$ 60  & 21 & 21     & 10" $\times$ 10"                    & 10\farcs3 $\times$ 9\farcs3   & 91    & 0.51 & 14.4 & 0.4 & 5\\
2014 May 11 & 2014.359    & 0.50 $\times$ 60   & 9 & 9       & 10" $\times$ 10"                    & 10\farcs1 $\times$ 9\farcs3   & 90    & 0.53  & 13.2 & 0.4 & 5\\
2014 Jul 2 & 2014.503        & 0.50 $\times$ 60   & 20 & 20     & 10" $\times$ 10"                    & 10\farcs2 $\times$ 9\farcs3   & 91    & 0.34 & 14.1 & 0.4 & 5\\
2014 Aug 4 & 2014.590      & 0.50 $\times$ 60    & 28 & 28     & 10" $\times$ 10"                    & 10\farcs2 $\times$ 9\farcs4   & 92    & 0.50 & 13.7 & 0.4 & 5\\
2016 May 21 & 2016.376   & 0.50 $\times$ 60    & 25 & 25     & 10" $\times$ 10"                     & 10\farcs6 $\times$ 9\farcs3 & 85 & 0.38 & 13.8 & 0.4 & 4\\
\enddata
\tablenotetext{a}{This is defined as the 95\% quantile of the distribution of magnitudes of detected stars with \textit{StarFinder.}}
\tablenotetext{b}{References: (0) Ghez et al. 2004; (1) Ghez et al. 2005a; (2) Hornstein et al. 2007; (3) Phifer et al. 2013;  (4) This work; (5) Witzel et al. 2014.}
\tablenotetext{c}{For our smaller field of view data ($\rm{L}^{\prime}$ in 20012 July and 2013 August and Ms), a PSF support size of 1" $\times$ 1" is used instead of our standard 2" $\times$ 2" support size in order to avoid edge effects. Comparisons to the larger field of view showed that his had no significant effect on the astrometry and photometry of G1.}
\end{deluxetable*}
\end{turnpage}

\section{Analysis}\label{analysis}
\subsection{Imaging Analysis}\label{imaganal}
Our imaging analysis is divided into two parts: (i) astrometric analysis using the PSF fitting tool \textit{StarFinder} and (ii) photometric and size calculations using a PSF convolved with a 2D elliptical Gaussian. Both measurements are described in detail below.

\begin{figure*}
\begin{center}
\includegraphics[width = 16.0cm]{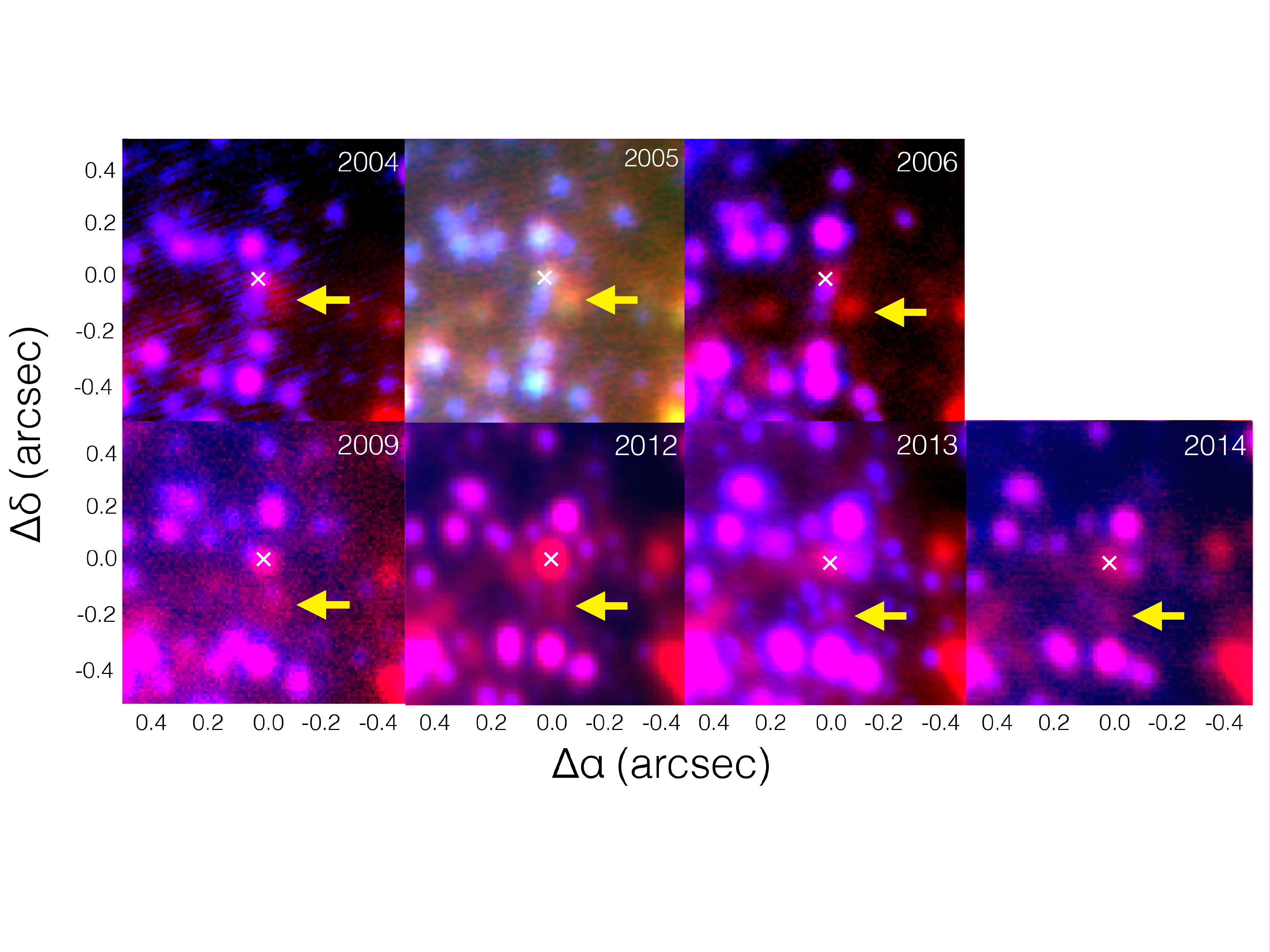}
\caption{Two-color images made by combining NIRC2 images at $\rm{K}^{\prime}$ (2.2 $\mu$m, in blue) and $\rm{L}^{\prime}$ (3.8 $\mu$m, in red). The images have been aligned using the coordinates of S0-2 from our respective \textit{StarFinder} runs for each filter during each epoch; the 2014 data are from our March 2014 observation. The position of Sgr A$^{*}$ is denoted by the white "$\times$", and the position of G1 is denoted by a yellow arrow. The 2005 panel shows a three-color image that includes our 2005 Ms data as well (2.2 $\mu$m in blue; 3.8 $\mu$m in green; 4.7 $\mu$m in red). G1 and G2 are distinctly red sources. Other red sources exist within the inner 0.5 arcseconds of Sgr A$^{*}$ as well and will be explored by Sitarski et al. (in preparation). It is apparent that Sgr A$^{*}$ varies considerably. For contour plots of G1, see Fig.~\ref{fig:g1cont}.}\label{fig:2col}
\end{center}
\end{figure*}

\subsubsection{Astrometry}\label{astrometry}
G1 is visually identified in every $\rm{L}^{\prime}$ and Ms image (see Fig.~\ref{fig:2col}). Its astrometric properties in every image were obtained using the PSF fitting program \textit{StarFinder} \citep{diolaiti2000} in a manner similar to what has been outlined in previous works (\citealt{yelda2014} and references therein). We initially ran \textit{StarFinder} using a correlation threshold of 0.8 and 0.6 in the main image and sub-images, respectively, to identify candidate sources in our images. This resulted in G1 detections in 2004, 2005, 2006, 2012, and 2013 in the $\rm{L}^{\prime}$ data. In 2003, 2009 and all 2014 $\rm{L}^{\prime}$ epochs and in both the Ms data sets, a different approach was necessary to capture G1 due to poorer data quality, although G1 can be visually identified (Fig.~\ref{fig:2col}). We therefore altered the search criterion to seek a point source within a three-pixel box centered at the point of the highest flux count, at the approximate position of G1 using a modified version of \textit{StarFinder} that searches for additional sources at a lower correlation \citep{boehle2014}. We do not use the 2016 $\rm{L}^{\prime}$ data for astrometry as we use the orbital model from 2003 - 2014 to predict the position of G1 in the 2016 data (see our photometric analysis described in Section~\ref{photsize}). With this modified procedure, G1 was detected in all epochs. 

While G1 is extended in the early epochs (Section~\ref{sizevar}), we still use the \textit{StarFinder} astrometry that reliably determines G1's centroid, as the residuals in each epoch look symmetric. Two-dimensional Gaussian fits convolved with a point spread function to G1 yielded peak positions consistent with the positions extracted from \textit{StarFinder}. 

The point sources identified in each epoch are matched across all epochs and transformed to a common coordinate system in which Sgr A$^{*}$ is at rest (see \citealt{phifer2013, yelda2010, yelda2014}; for the application of our distortion solution to L'-band data see Appendix~\ref{app1}.). Specifically, each $\rm{L}^{\prime}$ list of stellar positions (short: star list) is aligned to the $\rm{K}^{\prime}$ star list that is nearest in time with translation, rotation, and affine first-order transformation that is independent in $x$ and $y$. The transformed G1 position is added to the $\rm{K}^{\prime}$ star list and the $\rm{K}^{\prime}$ star lists from 1995 to 2014 are aligned as described in our earlier works (e.g., \citealt{ghez2008, yelda2014}) using measurements of infrared astrometric secondary standards taken through 2012 \citep{yelda2010, yelda2014, boehle2014}. Tab.~\ref{dataG1} lists the astrometry for G1 in each epoch prior to 2016.

\subsubsection{Photometry and Size Measurement}\label{photsize}
Magnitudes of all point sources at $\rm{K}^{\prime}$ were calculated using PSF fitting with \textit{StarFinder} procedure (see previous section). We chose IRS 16C, IRS 16NW, and IRS 16CC as our photometric calibrators, which is part of our standard $\rm{K}^{\prime}$ calibration procedure (e.g., \citealt{yelda2014}). 

As G1 seems extended in 2004, 2005, and 2006 at $\rm{L}^{\prime}$ and in the 2005 Ms data, we tested several photometric methods to obtain reliable photometry. To confirm whether the individual methods yielded reliable results during the epochs when G1 was visibly extended, we planted a 2D elliptical Gaussian model for the 2004 size (a Gaussian with a FWHM of $\sim58$~mas convolved with the PSF) in our data in three distinct regions: a high-background region, a low-background region, and near the position of G1. We planted sources with different brightnesses (mag$_{\rm{L}^{\prime}}$ = 10-16) to determine whether we could recover its magnitude and physical extent.

We tested three different photometric procedures: (1) \textit{StarFinder} with a PSF support array of 2.0 arcseconds (200 pixels), following our standard $\rm{K}^{\prime}$ analysis; (2) \textit{StarFinder} with a PSF support array of 0.9 arcseconds (90 pixels) to make the PSF more robust against background artifacts at larger distances from the core; (3)  a two dimensional fit with an intrinsic extended elliptical Gaussian source convolved with an empirical PSF model. The planting simulations returned significantly decreased fluxes with respect to their original planted magnitude in the case of \textit{StarFinder} PSF fitting with both PSF sizes $[$(1) and (2)$]$. However, (3) reliably recovered the fluxes and observed extent of the planted sources to within 20\% at the faintest magnitude tested (mag$_{\rm{L}^{\prime}}$ = 16).

We applied method (3) to every single $\rm{L}^{\prime}$ and Ms epoch (prior to 2016) using the IDL procedure \textit{mpfit2dfun}. To prepare the images for model fitting, we first subtracted all $\rm{L}^{\prime}$-detected \textit{StarFinder} sources. We then used the aligned $\rm{L}^{\prime}$ and $\rm{K}^{\prime}$ \textit{StarFinder} star lists to find the position of $\rm{K}^{\prime}$-only detected sources and used the forced \textit{StarFinder} algorithm from \citet{boehle2014} to find the fluxes of these sources at $\rm{L}^{\prime}$. We subtracted these sources as well as the \textit{StarFinder} generated backgrounds from the original image. In our 2D elliptical Gaussian model, we allowed the position angle to vary and allowed for the FWHM to range between 0.3 and 10.0 pixels. If the FWHM of G1 fell below 0.3 pixels ($\sim$3 mas) in an epoch, then a PSF without a Gaussian component was used instead. Our photometry is reported in Tab.~\ref{dataG1}. A comparison of the astrometry between the three methods yielded identical positions of G1 within 1$\sigma$ errors.

\begin{figure*}
\begin{center}
\includegraphics[width=16.0cm]{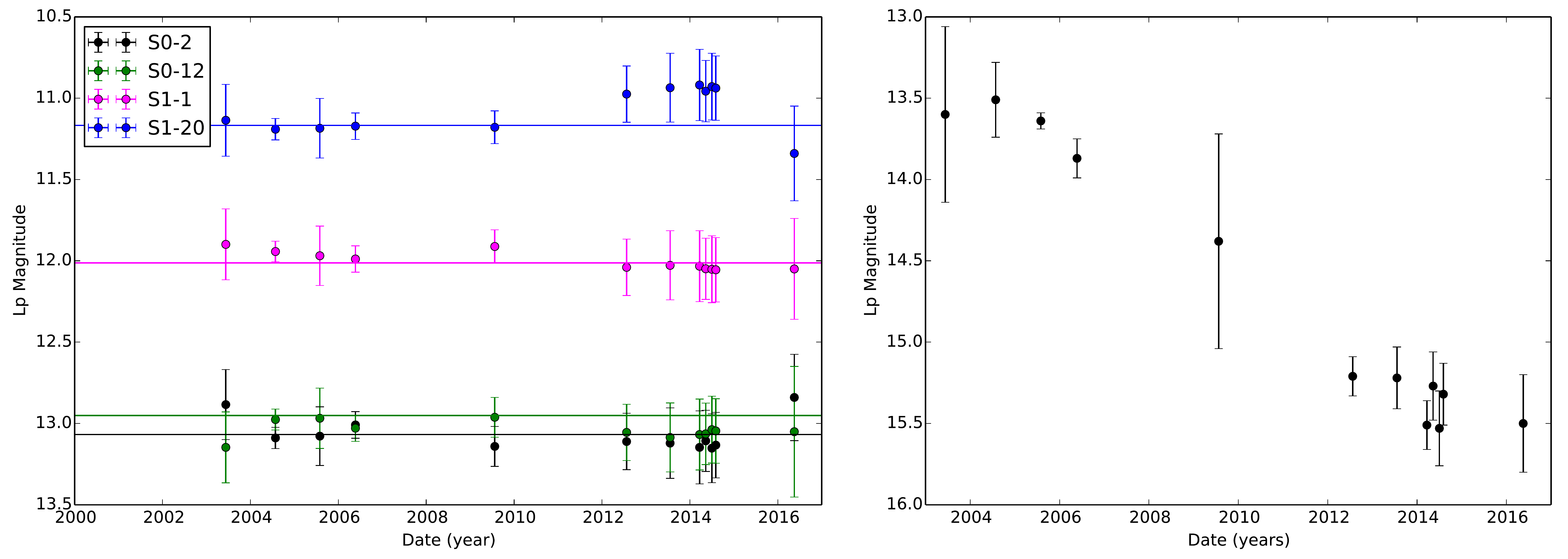}
\caption{\textit{Left}: $\rm{L}^{\prime}$ photometry on each of our four calibration sources. The solid line denotes the reported magnitude from \citet{schoedel2011}. \textit{Right}: $\rm{L}^{\prime}$ photometry of G1 a a function of time. The magnitude of the source has decreased significantly, and varies directly with the size of the source presented in Fig.~\ref{fig:size}.}\label{fig:phot}
\end{center}
\end{figure*}

To photometrically calibrate our $\rm{L}^{\prime}$ and Ms data, we used S0-2, S0-12, S1-20, and S1-1 and their $\rm{L}^{\prime}$ magnitudes from \citet{schoedel2010}. These sources were chosen because they are all in the field of view for every epoch, including our subarrayed epochs (see Tab.~\ref{obssum}). Similarly to \citet{schoedel2011}, the Ms data were calibrated using the same magnitudes as $\rm{L}^{\prime}$ because the relative colors of the calibrators are close to 0. The overall zero-point error from the \citet{schoedel2010} magnitudes is 0.15 magnitudes at $\rm{L}^{\prime}$ and is not taken into account in Tab.~\ref{dataG1} or Fig.~\ref{fig:phot} because they affect all photometric measurements in the same way.

No $\rm{K}^{\prime}$ counterpart was detected for G1 and star-planting simulations were performed to determine an upper magnitude limit. We used the $\rm{L}^{\prime}$ position of G1 in the 2013 August image, where G1 is an isolated point source (see Fig.~\ref{fig:2col}) and transformed that into the 2013 July $\rm{K}^{\prime}$ coordinate system. The star-planting simulations were carried out using our modified version of \textit{StarFinder} \citep{phifer2013, boehle2014}. There is a $\rm{K}^{\prime}$ source near G1 in 2013, S0-37, but it contributes at most 0.3 mJy to the overall $\rm{L}^{\prime}$ flux of G1 (assuming the dereddening law outlined in \citet{schoedel2010} and that S0-37 has the same colors as S0-2). All photometry in each bandpass is reported in Tab.~\ref{dataG1}. 

\begin{figure*}
\begin{center}
\includegraphics[width=16.0cm]{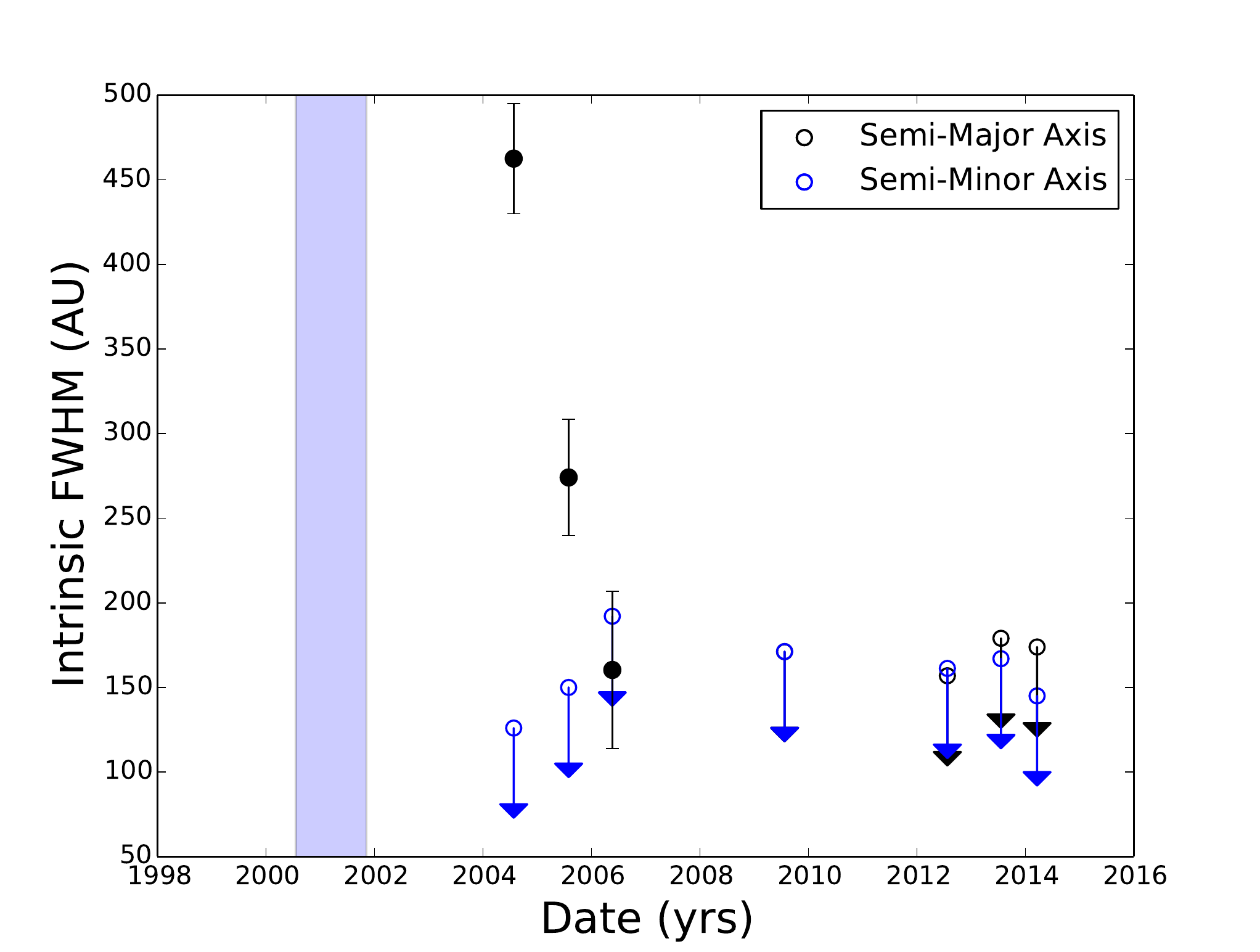}\caption{Size of G1 as a function of time since periapse passage ($T_{0} \sim$ 2001.3) in both the semi-major axis and semi-minor axis directions. In the epochs when G1 is resolved, we can get an actual measurement of the semi-major axis of the source; the last four epochs are upper limits on G1's size obtained by first subtracting out the closest neighboring sources in that epoch (S0-2 and Sgr A$^{*}$), and then comparing the 2-dimensional Gaussian profile of G1 to the point spread function. The light blue bar denotes the FWHM of the 1D marginalized probability distribution function for the periapse passage time.}\label{fig:size}
\end{center}
\end{figure*}

\begin{figure*}
\begin{center}
\includegraphics[width=12.0cm]{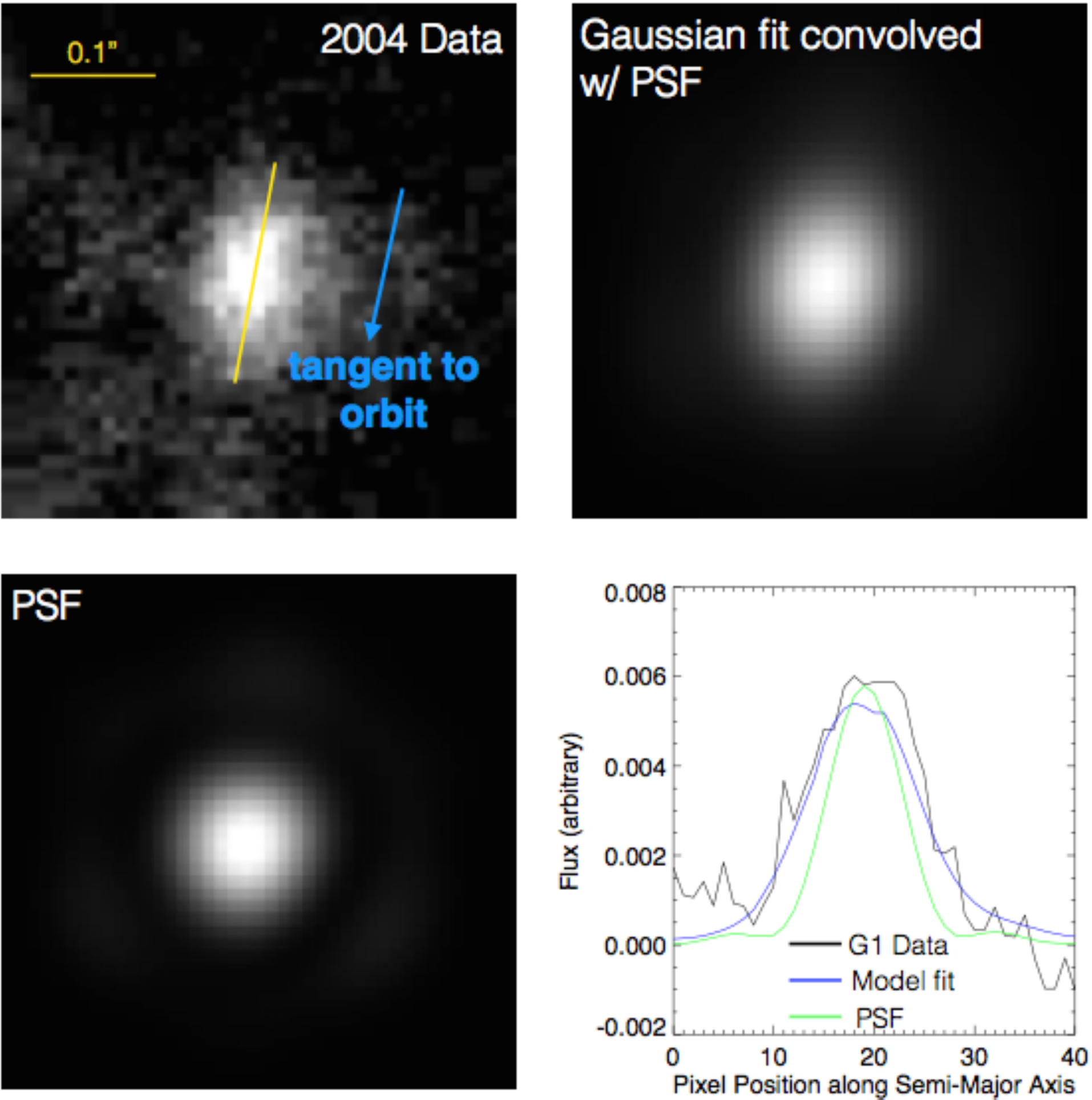}%gaussian_comparison_151120}
\caption{\textit{Upper Left}: G1 in 2004 after subtracting all $\rm{L}^{\prime}$ \textit{StarFinder}-detected point sources. The angle of the semi-major axis is denoted by the yellow line. The blue arrow shows a line in the direction of the tangent to the direction of orbital motion in 2004. \textit{Upper Right}: Image of 2-D Gaussian fit to the data found using \textit{mpfit2dpeak}. \textit{Lower Left}: PSF model from 2004 as extracted from \textit{StarFinder}. This panel, the upper left, and the upper right panel are all normalized so their peaks are on the same color table and scale; all panels are also on the same physical scale. \textit{Lower Right}: Slice along the semi-major axis for our data (black line), the 2-D Gaussian fit (blue line), and the PSF (green line). It is evident that the 2-D Gaussian fit is an acceptable model for the $\rm{L}^{\prime}$ extension and it is much larger than the PSF.}\label{fig:ext}
\end{center}
\end{figure*}

The recovered sizes of G1 from our model show that G1 is extended between 2004 and 2006, but is consistent with a point source from 2009 through 2014 (see Tab.~\ref{dataG1}). The magnitudes and sizes of G1 as a function of time are shown in Fig.~\ref{fig:phot} and Fig.~\ref{fig:size}, respectively, while Fig.~\ref{fig:ext} shows the elongation of G1 in the direction of orbital motion in 2004. The major axis angle of the 2D elliptical Gaussian is consistent with a tangential line to the orbit in 2004 and 2005 (10.4$\pm$4.0 degrees $[$tangent to orbit = 12.3$\pm$2.8$]$ and 27.4$\pm$4.8 degrees $[$tangent to orbit = 21.0$\pm$2.4$]$ west of north, respectively; see Fig.~\ref{fig:ext}).

\begin{deluxetable*}{lllllllllll}
\tabletypesize{\scriptsize}
\tablecaption{Data and Observed Properties of G1 \label{dataG1}}
\tablewidth{0pt}
\tablehead{\colhead{Date} & \colhead{$\Delta$RA\tablenotemark{a}} & \colhead{$\Delta$Dec\tablenotemark{a}} & \colhead{$\rm{K}^{\prime}$} & \colhead{$\rm{L}^{\prime}$} & \colhead{Ms} & \colhead{$\rm{L}^{\prime}$ Semi-Major Axis} & \colhead{PA of Gaussian}\\
 & \colhead{(arcsec)} & \colhead{(arcsec)}& \colhead{(mag)} & \colhead{(mag)} & \colhead{(mag)}& \colhead{Intrinsic Size (AU)} & \colhead{2D fit (deg)}}
\startdata
2003.44 & -0.077$\pm$0.009 & -0.059$\pm$0.009 & & 13.60$\pm$0.54 & & & \\
2004.57 & -0.073$\pm$0.009 & -0.068$\pm$0.008 & & 13.51$\pm$0.23 & & 460$\pm$30 & 10.4$\pm$4.0\\
2005.58 & -0.069$\pm$0.007 & -0.860$\pm$0.006 & & 13.64$\pm$0.05 & 12.71$\pm$0.30 & 270$\pm$30 & 27.4$\pm$2.4\\
2006.39 & -0.065$\pm$0.006 & -0.103$\pm$0.006 & & 13.87$\pm$0.12 & & 160$\pm$50&16.0$\pm$9.8\\
2009.56 & -0.035$\pm$0.006 & -0.131$\pm$0.010 & & 14.38$\pm$0.66 & & $<$170&\\
2012.56 & -0.029$\pm$0.006 & -0.162$\pm$0.007 & & 15.21$\pm$0.12 & & $<$160&\\
2013.55 & -0.012$\pm$0.006 & -0.169$\pm$0.006 & $>$18.8 & 15.22$\pm$0.19 & & $<$180&\\
2014.22 &  0.002$\pm$0.006 & -0.183$\pm$0.009 & & 15.51$\pm$0.15 & & $<$170&\\
2014.36 &  0.003$\pm$0.006 & -0.196$\pm$0.006 & & 15.27$\pm$0.21 & & &\\
2014.50 &  0.003$\pm$0.006 & -0.193$\pm$0.007 & & 15.53$\pm$0.23 & & &\\
2014.59 &  0.006$\pm$0.006 & -0.185$\pm$0.007 & & 15.32$\pm$0.19 & & &\\
2016.38 &  0.017$\pm$0.004\tablenotemark{b} & -0.202$\pm$0.004\tablenotemark{b} & $>$18.2\tablenotemark{c} & 15.50$\pm$0.36\tablenotemark{c} & 14.81$\pm$0.23\tablenotemark{c} & & 
\enddata
\tablenotetext{a}{{$\Delta$RA and $\Delta$Dec are given in a reference frame in which the location of the black hole is not at the origin (see Tab.~\ref{fitBHpar}).}}
\tablenotetext{b}{These positions come from our orbital solution derived from our $\rm{L}^{\prime}$ data taken from 2003 through 2014. See Section~\ref{orbg1} for more information on our orbital fit.}
\tablenotetext{c}{The photometry in 2016 was derived using deconvolved images rather than \textit{StarFinder}, as done in previous epochs. See Section~\ref{photsize} for more information.}
\end{deluxetable*}

In order to be able to infer the blackbody properties of G1 in a later epoch when the source is compact, we utilize $\rm{L}^{\prime}$ and Ms data from 2016. As G1 is in a confused region in this epoch, we adopt a different methodology to recover its $\rm{K}^{\prime}$ upper limit and $\rm{L}^{\prime}$ and Ms flux densities. The \textit{StarFinder}-generated backgrounds are subtracted from each main map and we subtract all \textit{StarFinder} identified point sources in the vicinity of our predicted position for G1. We then use the \textit{StarFinder}-generated PSF to do an iterative Lucy-Richardson deconvolution (with 10,000 iterations). Aperture photometry is then performed after beam restoring at the predicted position of G1 based on its derived orbit from the 2003 - 2014 data. The photometry of the 2016 $\rm{L}^{\prime}$ data matches well with the 2014 epochs.

\subsection{Spectroscopic Analysis}\label{specanal}
\begin{figure*}
\begin{center}
\includegraphics[width=18cm]{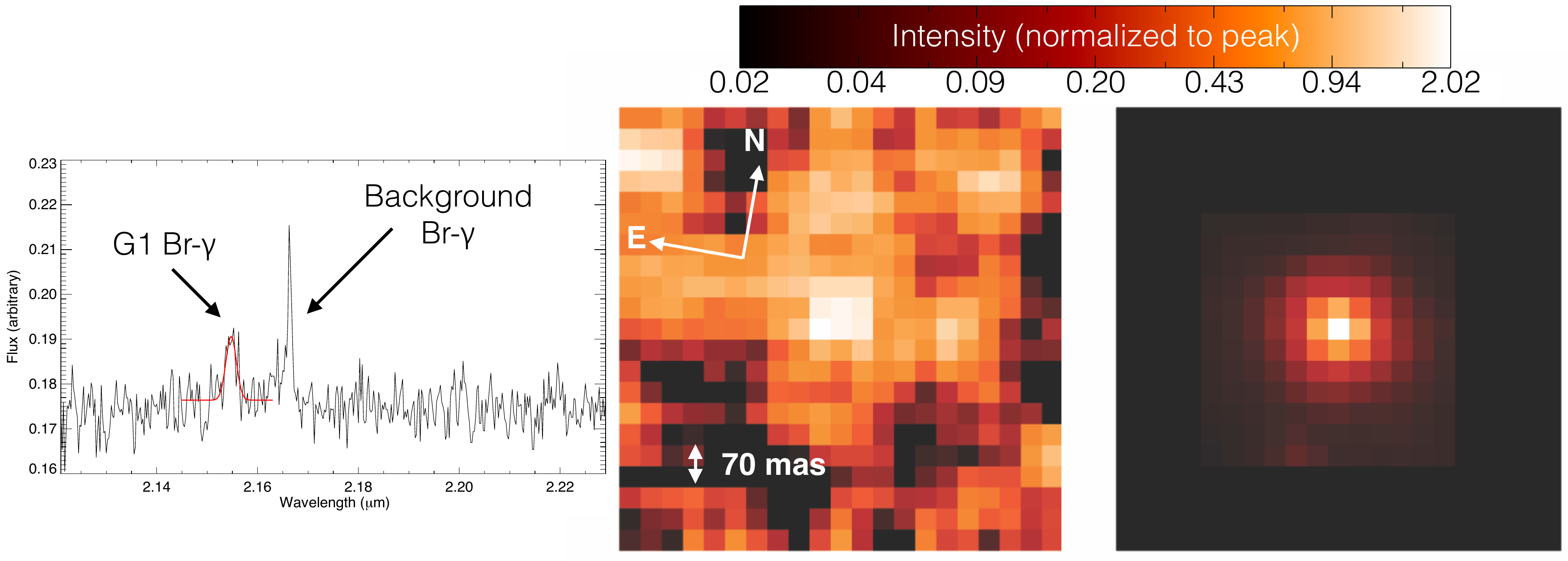}%{spec_cube_figure_151012}
\caption{\textit{Left}: Continuum-subtracted spectrum of G1. The emission line at 2.154 $\mu$m corresponds to a radial velocity of -1568 $\pm$ 60 km/sec. This spectrum was extracted using an aperture of 1 pixel (35 mas) radius from our 2006 OSIRIS data cube. The overplotted red line shows the 1-dimensional Gaussian fit.  \textit{Center}: Continuum-subtracted OSIRIS data cube collapsed over a $\Delta$v of 267 km/sec centered on -1568 km/sec (LSR-corrected) and smoothed over 2 spatial pixels. \textit{Right}: Br-$\gamma$ point spread function extracted from our 2006 OSIRIS data cube. Both the PSF and the collapsed cube are displayed on the same physical scale and same logarithmic color scale where we normalize each figure to its respective peak.}\label{fig:spec}
\end{center}
\end{figure*}

The radial velocity for G1 was obtained using a similar approach to that developed by \citet{phifer2013} for G2: Aa spectrum of G1 was extracted from our 2006 OSIRIS data at a location that was determined by transforming the high quality 2005 $\rm{L}^{\prime}$ star list to the OSIRIS coordinate system using a second-order polynomial transformation. The spectrum was extracted using an aperture radius of 1 pixel (corresponding to 35 mas) and applying a local sky subtraction in an area clear of known contaminating stars (e.g., \citealt{do2013b}). The extracted spectrum was calibrated using the standard techniques \citep{do2009a}, and the peak in the resulting spectrum was fit with a Gaussian model to derive an observed radial velocity and its full-width at half-maximum. The resulting heliocentric radial velocity was corrected by 3.64 km sec$^{-1}$ to correspond to an LSR velocity of -1568 $\pm$ 60 km sec$^{-1}$ on the date of the observation (see Tab.~\ref{radvel}). The FWHM of the spectral line is 185 $\pm$ 41 km sec$^{-1}$. The spectrum and the corresponding point source in the line emission map are shown in Fig.~\ref{fig:spec}. 

\begin{deluxetable}{llll}
\tabletypesize{\scriptsize}
\tablecaption{Radial Velocity Data \label{radvel}}
\tablewidth{0pt}
\tablehead{\colhead{Date} & \colhead{Radial Velocity} & \colhead{PSF FWHM} & \colhead{Orig. Publication}\\
& \colhead{km/sec} & \colhead{mas at Br-$\gamma$} & }
\startdata
2004.6 & -2043$\pm$150 & & \citealt{pfuhl2015}\\
2006.2 & -1594$\pm$163 & & \citealt{pfuhl2015}\\
2006.5 & -1558$\pm$60 & 67 & This Paper\\
2008.3 & -1123$\pm$159 & & \citealt{pfuhl2015}
\enddata
\end{deluxetable}

We compute the Br-$\gamma$ line luminosity similarly to \citet{phifer2013} by comparing S0-2's flux density to G1's flux density. We estimate S0-2's dereddened flux density to be 14.1 $\pm$ 0.2 mJy (assuming the extinction prescription outlined in \citealt{schoedel2010} and the flux density from \citealt{ghez2008}) and compute an expected luminosity of S0-2 over the 2.15 - 2.159 $\mu$m bandpass to be 0.16$\Lsun$. We then integrate over the same bandpass on the S0-2 and G1 spectra to get a final Br-$\gamma$ luminosity of G1 of $(1.48 \pm 0.17) \cdot 10^{-3}~\Lsun$. To check for consistency, we followed this same procedure to integrate over the same bandwidth for G2 (2.17 - 2.179 $\mu$m), which yields a Br-$\gamma$ luminosity of $(1.36 \pm 0.25) \cdot 10^{-3}~\Lsun$, consistent within 1$\sigma$ of the 2006 G2 luminosity reported in \citet{phifer2013}.

Using this dereddened Br-$\gamma$ line luminosity from 2006 (in an epoch where G1 is resolved), we can estimate the Lyman-$\alpha$ emission luminosity. We used the relationship between the Br-$\gamma$ emission and the free-free emission given in \citet{wynnwilliams1978} and solved for the Lyman-$\alpha$ luminosity using the formulae summarized in \citet{genzel1982} and \citet{becklin1993}. We estimate that the Lyman-$\alpha$ luminosity is $\sim$2 $\Lsun$.

\subsection{The Orbital Determination of G1}\label{orbg1}
To derive the orbital properties of G1, we jointly fit for the Keplerian orbital parameters of S0-2, S0-38, and G1 (period, epoch of periapse passage, eccentricity, position angle of the ascending node, argument of periapse, and inclination for each source) and the black hole parameters (the two-dimensional position, the three-dimensional velocity, and the mass of and distance to Sgr A$^{*}$; see Tab.~\ref{fitBHpar}).  S0-2 and S0-38 have complete orbital phase coverage and drive the black hole parameter fit. We use the same astrometry and radial velocities of S0-2 and S0-38 as reported in \citet{boehle2014}. Jointly fitting the three sources enables us to find an orbital solution for G1 and for the black hole parameters. G1 alone does not have enough kinematic information to independently fit for the black hole parameters due to the lack of orbital phase coverage. We impose uniform, flat priors on each of the orbital parameters for G1 as follows: $[$0, 360$]$ degrees for the argument of periapse ($\omega$); $[$0, 180$]$ degrees for the position angle of the ascending node ($\Omega$); $[$0, 180$]$ degrees for the inclination; $[$0, 1$]$ for the eccentricity; $[$0, 6000$]$ years for the period; and $[$1990, 2010$]$ for the time of periapse passage. The G1 astrometry consists of 11 data points (see Section~\ref{imaganal}), including our newest astrometric measurements. For this orbital fit, we also used the three radial velocity measurements reported in \citet{pfuhl2015} and our new measurement (Section~\ref{specanal}; Tab.~\ref{radvel}). 

In this study, we include three sources of hypothetical systematics uncertainties: (i) source confusion, (ii) the presence of outliers and (iii) uncertainty arising in the construction of the absolute reference frame. 

Source confusion is a significant source of systematic error in our orbital fits and the formal uncertainties are therefore underestimated. In order to account for this, we fit a second-order polynomial to our astrometric data points and add a single additive value in quadrature to the formal errors until the final reduced chi-squared of the second-order fit (that includes position, velocity, and acceleration) is equal to 1.0. The resulting additive value is 5.5 mas which is roughly comparable to the formal uncertainties.

In order to assess the impact of hypothetical outliers, we use a Jackknife resampling method where we fit for G1's orbital parameters by dropping one epoch of observations at a time. This analysis, presented in Appendix~\ref{jackg1}, shows results that are consistent with the ones reported in Tab.~\ref{fitBHpar} and we conclude that no outlier impacts significantly our results.

{Following the procedure described in the Appendix of \citet{boehle2014}, the systematic uncertainties due to the construction of the absolute reference frame have been assessed by using a Jackknife resampling method. In this Jackknife analysis, one of the seven masers used to construct the reference frame is dropped at a time. \citet{boehle2014} showed that this systematic uncertainty is important for the SMBH position and velocity relative to the origin of the constructed reference frame, but is negligible for all the other fitted parameters such as the SMBH mass and distance $R_0$ and the orbital parameters. We conclude that systematic effects arising from the construction of our absolute reference frame do not impact significantly the posterior probability distribution for stellar orbital parameters.}
\begin{deluxetable}{lllllllllll}
\tabletypesize{\scriptsize}
\tablecaption{S0-2 + S0-38 Black Hole Parameter Values \label{fitBHpar}}
\tablewidth{0pt}
\tablehead{\colhead{Orbital Parameter} & \colhead{Peak Fit\tablenotemark{a}}}
\startdata
X-Position of Sgr A* ($x_{0}$, mas) & 2.1$^{+0.5}_{-0.3}$$\pm$1.90\\
Y-Position of Sgr A* ($y_{0}$, mas) & -4.4$\pm$0.4$\pm$1.23\\
$\Delta$RA Velocity of Sgr A* ($V_{x}$, mas/yr) & -0.12$^{+0.03}_{-0.02}$$\pm$0.13\\
$\Delta$Dec Velocity of Sgr A* ($V_{y}$, mas/yr) & 0.68$\pm$0.05$\pm$0.22\\
Radial Velocity of Sgr A*($V_{z}$, km/sec) & -20.4$\pm$6.3$\pm$4.28\\
Distance to Sgr A* ($R_{0}$, kpc) & 7.87$\pm$0.11\\
Mass of Sgr A* ($M$, Millions of $\Msun$) & 3.93$^{+0.07}_{-0.13}$\\
\hline
\hline
S0-2 Parameters:\\
\hline
\hline
Time of closest approach ($T_{0}$, years) & 2002.346$\pm$0.003\\
Eccentricity ($e$) & 0.892$^{+0.002}_{-0.001}$\\
Period ($P$, years) & 15.93$^{+0.02}_{-0.05}$\\
Angle to periapse ($\omega$, degrees) & 66.8$^{+0.3}_{-0.5}$\\
Inclination ($i$, degrees) & 134.3$\pm$0.3\\
Position angle of the ascending node ($\Omega$, degrees) & 228.0$^{+0.4}_{-0.5}$\\
\hline
\hline
S0-38 Parameters:\\
\hline
\hline
Time of closest approach ($T_{0}$, years) & 2003.191$^{+0.038}_{-0.017}$\\
Eccentricity ($e$) & 0.811$\pm$0.003\\
Period ($P$, years) & 19.22$^{+0.1}_{-0.2}$\\
Angle to periapse ($\omega$, degrees) & 13$^{+15}_{-21}$\\
Inclination ($i$, degrees) & 169$\pm$2\\
Position angle of the ascending node ($\Omega$, degrees) & 94$^{+18}_{-14}$
\enddata
\tablenotetext{a}{The peak and corresponding 1$\sigma$ errors are from the marginalized one-dimensional distributions for the respective parameters. The first error term corresponds to the error determined by the orbital fit itself, while the second error term on the black hole parameters refers to uncertainty in the reference frame and was determined by \citet{boehle2014}.}
\end{deluxetable}

\begin{figure*}
\begin{center}
\includegraphics[width=16cm]{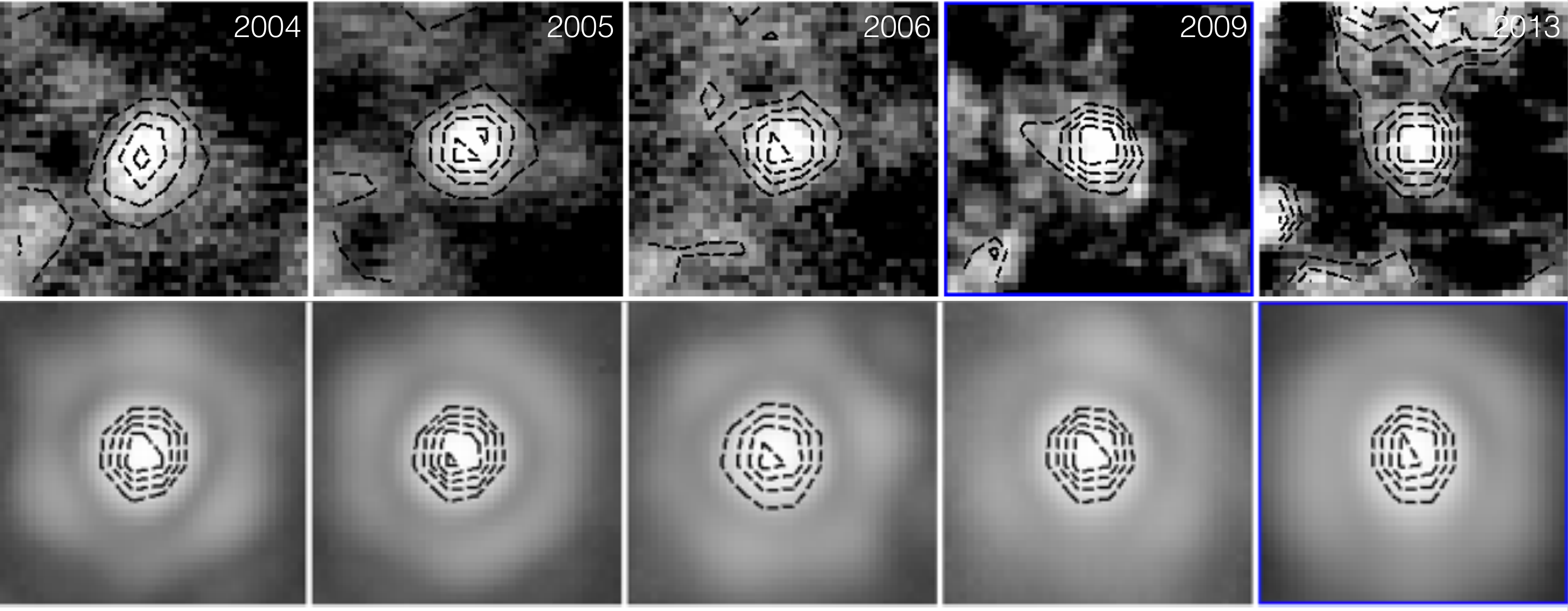}
\caption{\textit{Top}: Source-subtracted images centered on G1 that are 0.4 arcseconds on a side. Each image is photometrically normalized to a constant flux. \textit{Bottom}: Point spread functions corresponding to the epochs in the top row. The contours show intensity levels on the same levels as those in the top row. G1 is extended in 2004 - 2006, whereas G1 is compact after 2009.}\label{fig:g1cont}

\end{center}
\end{figure*}

\begin{figure*}
\begin{center}
\includegraphics[width = 16.0cm]{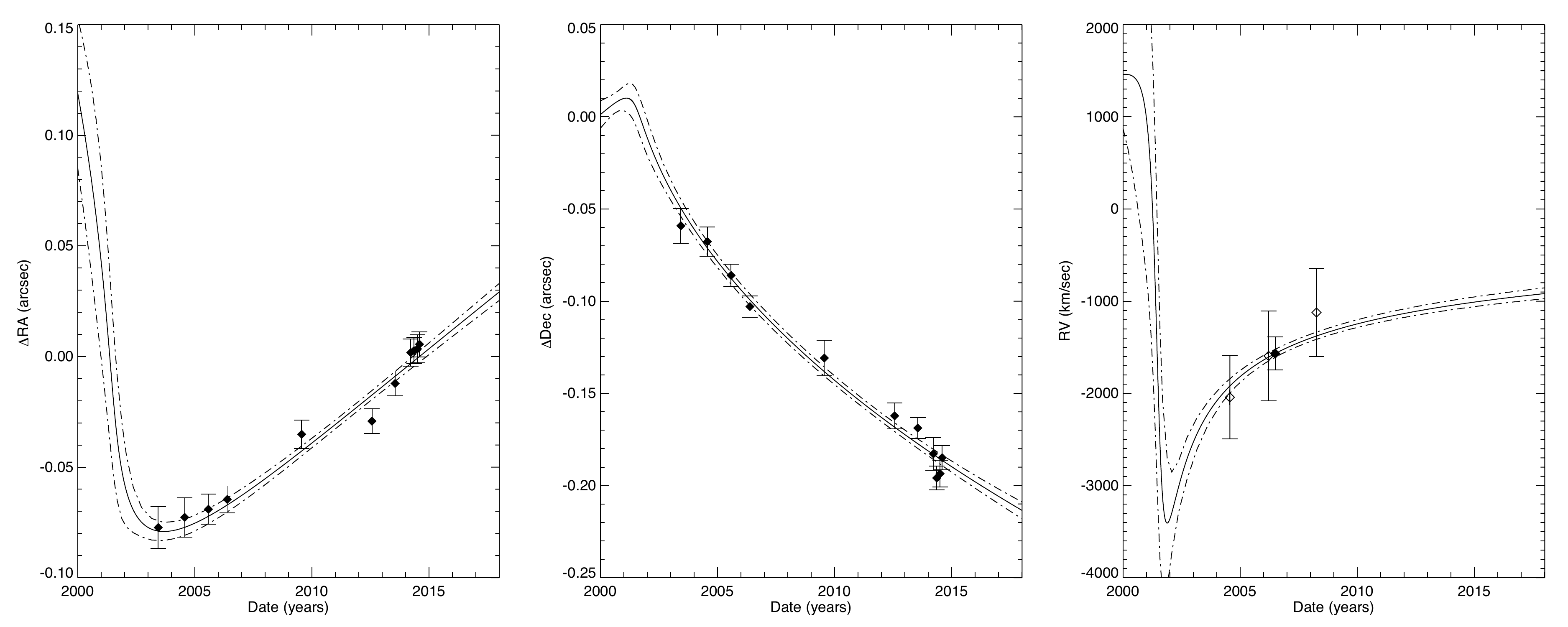}
\caption{G1's kinematic measurements and best fit orbital motion models. Our 11 astrometric and 1 radial velocity measurements are shown as filled points. The three RV measurements from \citet{pfuhl2015} are plotted as unfilled points. The 1$\sigma$ uncertainty envelopes are shown as broken lines.}
\label{fig:best_fit_model}
\end{center}
\end{figure*}

\begin{figure*}
\begin{center}
\includegraphics[width=16.cm]{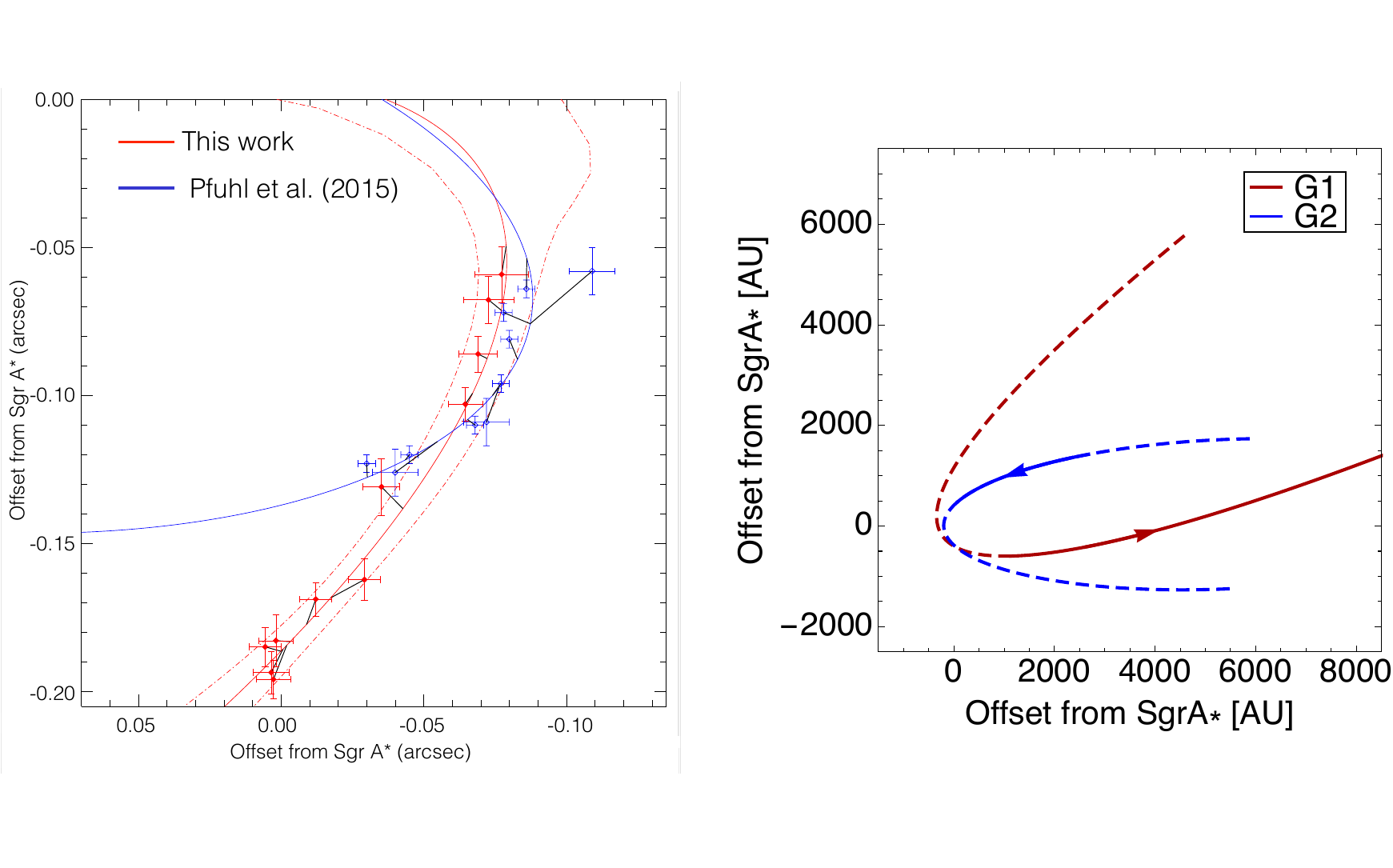}
\caption{\textit{Left:} Comparison of G1's orbital solution between this work and \citet{pfuhl2015}. Our Keplerian orbital fit is shown in red, while the orbital fit and data from \citet{pfuhl2015} are shown in blue. The black lines connect the observed point to the same point in time on the model orbit. There is an astrometric bias in 2009 and 2010 from confusion or a background dust emission feature that may skew the astrometry in those epochs. We do not use the astrometry from \citet{pfuhl2015} due to differing reference frames. \textit{Right:} Orbits of G1 and G2 (as described by Table~\ref{tab:orbital_solution}) projected into their common average orbital plane ($\Omega_{\rm{G1}}$ = +2.5 deg; $\Omega_{\rm{G2}}$ = -2.5 deg; $i_{\rm{G1}}$ = -2 deg; $i_{\rm{G2}}$ = 2 deg; $\omega_{\rm{G1}}$ = 117 deg; $\omega_{\rm{G2}}$ = 96 deg). It is evident that despite having similar orbital planes, the orbital trajectories are different. The solid (dotted) lines show times when we have (have not) observed G1 and G2.}
\label{fig:comp_orbit}
%\textit{Right:} Projected distance from Sgr A$^{*}$ as a function of time. The astrometric points reported in this work are over-plotted on our best-fit orbital solution.}
\end{center}
\end{figure*}

\begin{figure*}
\begin{center}
\includegraphics[width=17.0cm]{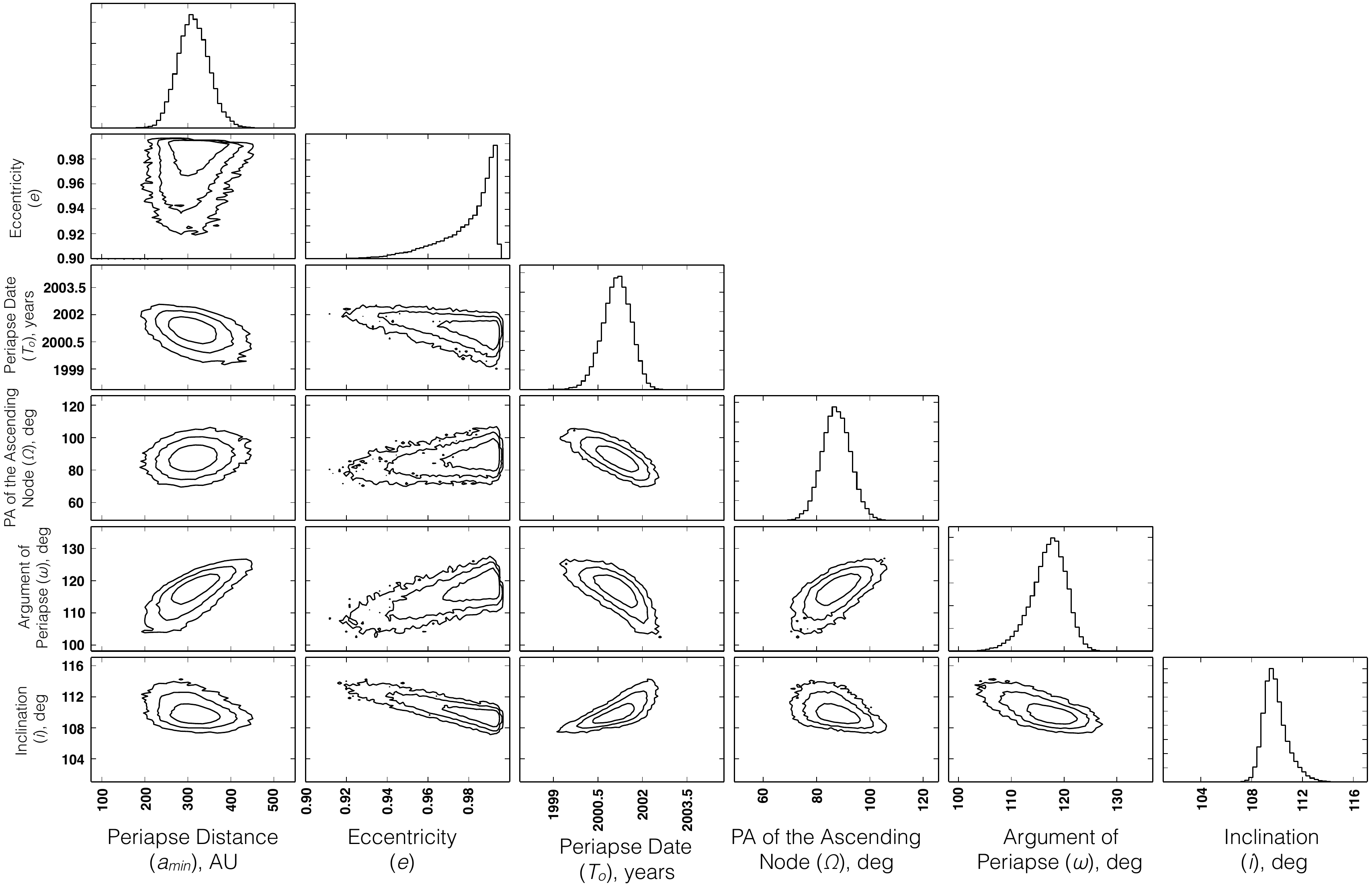}%{triangle_plot_G1only_150925}
\caption{One-dimensional marginalized probability distribution functions for the six Keplerian orbital parameters for G1 (black 1, 2, and 3$\sigma$ contours), along with the joint probability distribution functions for all parameters.}
\label{fig:corner}
\end{center}
\end{figure*}

\section{Results}\label{results}
Our analysis of both photometric and spectroscopic information and our Keplerian orbital fit have led to three key results: G1 follows a highly eccentric Keplerian orbit that differs from G2's orbit; shortly after periapse, G1's $\rm{L}^{\prime}$ emission is extended along the direction of orbital motion; and G1's $\rm{L}^{\prime}$ emission is much larger than the tidal radius of even a 100$\Msun$ source shortly after periapse, indicating that this emission comes from material that is not gravitationally bound to G1.

\subsection{Keplerian Orbital Fit Results}
\begin{figure*}
\begin{center}
\includegraphics[width=18.0cm]{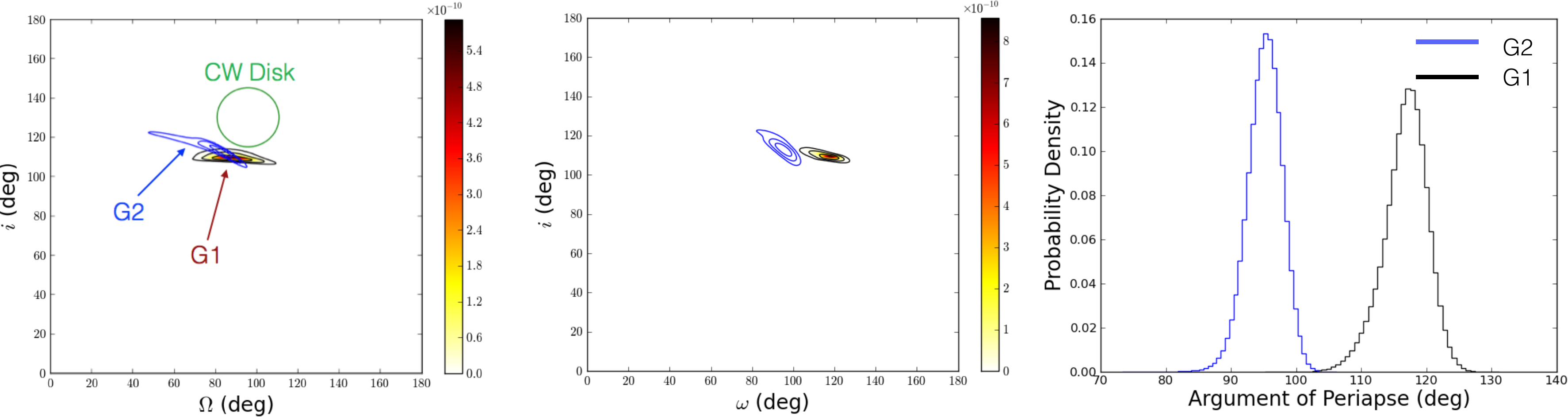}
\caption{Joint probability distribution functions showing G1 (black) and G2 (blue) 3$\sigma$ contours. While G1 and G2 may have similar orbital orientations, as shown in the $\Omega$ vs. $i$ plot (\textit{left}), their arguments of periapse ($\omega$) differ by greater than 3$\sigma$, implying that they have different orbits (\textit{center, right}). The clockwise disk's orientation and width are overplotted on the $\Omega$ vs. $i$ plot to show the orientation of the orbital plane's proximity to the clockwise-moving disk of young stars.}
\label{fig:2d_posterior}
\end{center}
\end{figure*}

The orbit of G1 is consistent with Keplerian motion (see Figures~\ref{fig:best_fit_model} and~\ref{fig:comp_orbit}). Based on our precise astrometry and radial velocity points, G1 lies on a highly eccentric orbit ($e$ = 0.99$^{+0.001}_{-0.01}$) and has recently passed through periapse ($T_{0}$ = 2001.3 $\pm$ 0.4). The three orbital angles (position angle of the ascending node $\Omega =  87.1^{+5.0}_{-4.9}$, argument of periapse $\omega = 117.3^{+2.8}_{-2.9}$, and inclination $i =109.0^{+0.9}_{-0.8}$) are well-constrained, but the orbital period is very poorly constrained due to lack of orbital phase coverage. {The 1D and 2D joint posterior distribution function are shown in Figures~\ref{fig:corner} and~\ref{fig:2d_posterior}.} The best-fit orbit is shown in Fig.~\ref{fig:comp_orbit} and the peak of the 1-dimensional marginalized probability distribution functions along with the maximum likelihood best fit are presented in Table~\ref{tab:orbital_solution}. We only fit bound, closed orbits; G1 could be on a hyperbolic orbit since the eccentricity distribution is artificially truncated. The period restriction of $P<$6000 years also constrains our orbital fits. 

Fig.~\ref{fig:best_fit_model} shows our orbital plots, our extracted G1 astrometry and radial velocity measurements. Fig.~\ref{fig:comp_orbit} presents our projection of the orbit onto the sky and compares it to the orbital solution from \citet{pfuhl2015} while assuming the black hole parameters from \citet{gillessen2009}. {The left of Fig.~\ref{fig:comp_orbit} shows clearly that our optimal G1 solution is significantly different from the previously published solution from \citet{pfuhl2015}. As a consequence, the best-fit G1 orbit is no longer in agreement with G2's optimal solution (right panel of Fig.~\ref{fig:comp_orbit}; G2's orbit is thoroughly discussed in \citealt{gillessen2012, gillessen2013b, phifer2013}, and \citealt{meyer2013}). This can be understood because our data covers almost twice the time baseline presented in \citet{pfuhl2015}.}

\begin{deluxetable*}{lllllllllll}
\tabletypesize{\scriptsize}
\tablecaption{Orbital Parameters for G1 and G2 \label{tab:orbital_solution}}
\tablewidth{0pt}
\tablehead{\colhead{Parameter} & \colhead{Best Fit, G1} & \colhead{Peak, G1\tablenotemark{b}} & \colhead{Best Fit, G2\tablenotemark{c}} & \colhead{Peak, G2\tablenotemark{c}} & \colhead{G1 Fit}\\
& & & & & \citet{pfuhl2015}}
\startdata
Time of closest approach ($T_{0}$, years) & 2001.0 & 2001.3$^{+0.4}_{-0.2}$ & 2014.1 & 2014.2$^{+0.03}_{-0.05}$ & 2001.6$\pm$0.1\\
Eccentricity ($e$) & 0.981 & 0.992$^{+0.002}_{-0.01}$ & 0.962 & 0.964$^{+0.036}_{-0.073}$ & 0.860$\pm$0.050\\
Periapse Distance ($A_{\rm{min}}$, AU) & 277 & 298$^{+32}_{-24}$ & 193 & 201 $\pm$ 13 & 417 $\pm$ 239 \\
Argument of periapse ($\omega$, degrees) & 118 & 117$\pm$3 & 95 & 96$\pm$2 & 109$\pm$8\\
Inclination ($i$, degrees) & 109 & 109$\pm$1 & 112 & 113$\pm$2 & 108$\pm$2\\
Position angle of the ascending node ($\Omega$, degrees) & 89 & 88$^{+5}_{-4}$ & 83 & 82$\pm$2 & 69$\pm$5

\enddata
\tablenotetext{a}{The parameters of \Sgra are extracted as described above.}
\tablenotetext{b}{The errors reported here are the 1$\sigma$ errors taken from the marginalized one-dimensional distributions for the respective parameters.}
\tablenotetext{c}{G2 parameters are from performing an orbital fit on our available astrometric and spectroscopic points (those outlined in \citealt{meyer2013}) in the same fashion described in Section~\ref{orbg1}.}
\tablenotetext{}{The clockwise disk parameters are $i$=130$\pm$15 deg and $\Omega$=96$\pm$15 deg, where 15 deg reflects the half-width at half-maximum from the peak density of the clockwise disk as reported in \citealt{yelda2014}.}
\end{deluxetable*}

\subsection{Size Variation}\label{sizevar}
In the epochs closest to periapse, G1 is extended along the direction of orbital motion (the semi-major axis). Fig.~\ref{fig:size} shows the intrinsic extent of G1 corrected for the size of the PSF along both the semi-major and semi-minor axes. The source shape is approximately elliptic and the semi-major axis of an elliptical 2D Gaussian fit is aligned with the direction of linear motion (Fig.~\ref{fig:ext}). However, in the more recent epochs, G1 becomes more compact. Additionally, there is significant brightness variation of G1 at $\rm{L}^{\prime}$ post-periapse passage, which corresponds to its size evolution: when G1 is at its largest size, it is also brightest; when G1 is compact, it is $\sim$2 magnitudes dimmer. The arrows in Fig.~\ref{fig:size} show the intrinsic (PSF-size corrected) upper limits on the source size along the semi-major and semi-minor axes, which is on average $\sim$170 AU along the semi-major axis assuming $R_{0} = 8$ kpc.

\begin{figure*}
\begin{center}
\includegraphics[width=15.0cm]{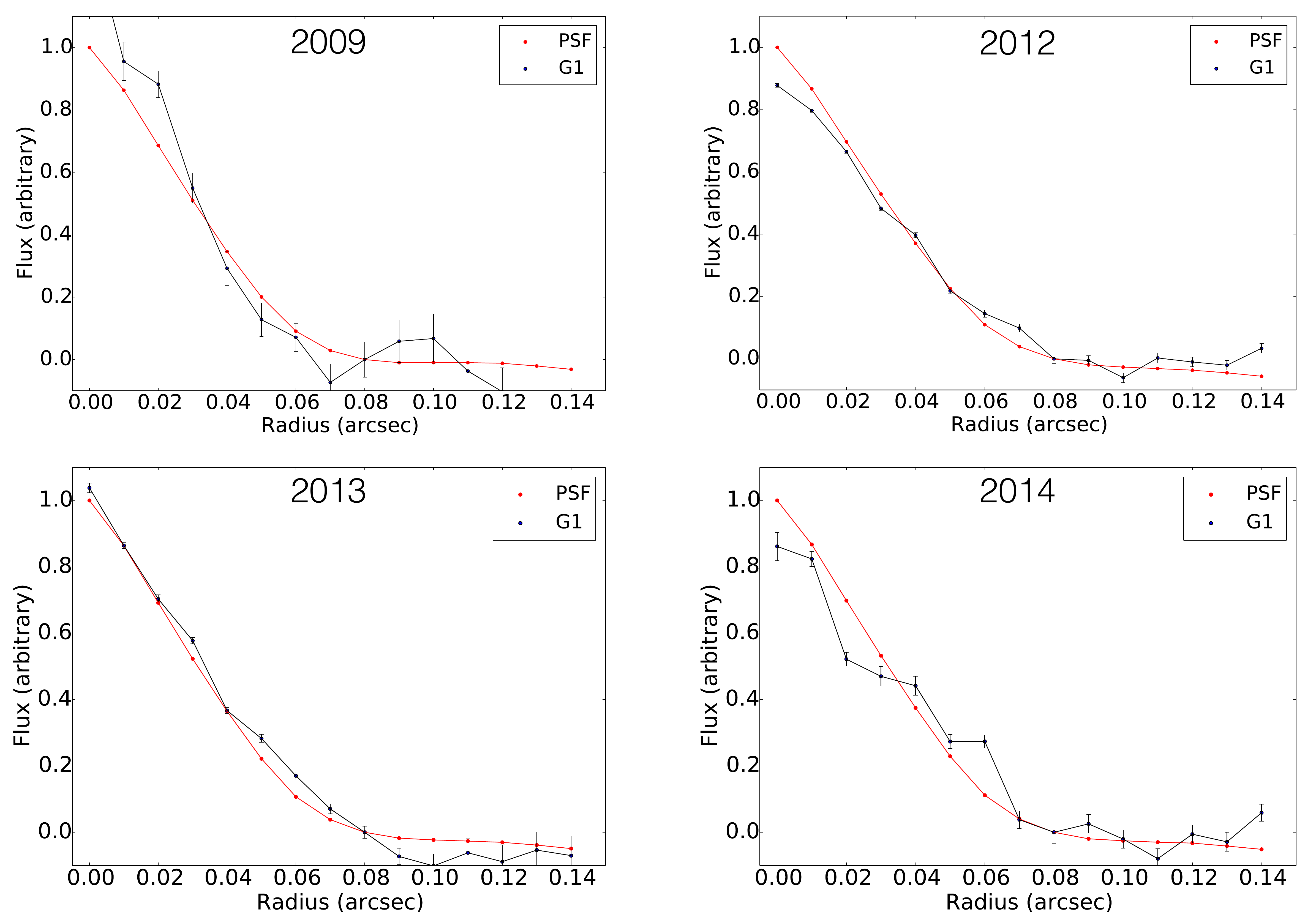}
\caption{Azimuthally-averaged radial profiles for G1 from 2009 through 2014. It is evident that the size of G1 becomes consistent with our \textit{StarFinder}-extracted model of our point spread function.}\label{fig:azi}
\end{center}
\end{figure*}

Fig.~\ref{fig:g1cont} shows images of G1 with all neighboring point sources identified by \textit{StarFinder} subtracted. The contours illustrate the size development of G1. The full-width at half-maximum of the semi-major axis of G1 is as high as 463 $\pm$ 16 AU in 2004.567 after correcting for the PSF contribution (see Fig.~\ref{fig:size}), but decreases to the size of a point source after 2006. Fig.~\ref{fig:azi} also shows azimuthally-averaged radial profiles of G1 from 2009 through 2014, showing that the size of G1 is indeed consistent with a point source.

Our 2006 Br-$\gamma$ detection is quite shallow and we are unable to determine whether G1 is resolved at Br-$\gamma$. Due to the shallowness of the Br-$\gamma$ detection, we are unable to conclude if G1 is spatially resolved or has a velocity gradient.

\subsection{Photometry and Temperature of G1}\label{phottemp}
There is a large photometric difference ($\sim$2 magnitudes) between the epochs when G1 is extended (2004, 2005, and 2006) and when it is point-like. The brightness develops with size, as epochs when G1 is extended are brightest, and epochs when G1 is point-like are dimmer and remain at a constant magnitude from 2012 through 2016.

G1 is identified at $\rm{L}^{\prime}$ and Ms ($\rm{L}^{\prime}$ = 13.65 in 2005; Ms = 12.71 in 2005), but not at $\rm{K}^{\prime}$ ($\rm{K}^{\prime}$ $>$ 18.8 in 2013). Assuming zero-point fluxes for $\rm{L}^{\prime}$ from \citet{tokunaga2000} and the extinction law outlined in \citet{schoedel2010}, we infer a dereddened $\rm{L}^{\prime}$ flux of 2.7 $\pm$ 0.5 mJy in 2005.

In order to infer a temperature for G1 at a moment in time (2005) when G1 is extended enough to be resolved, we expect it to be optically thin, and therefore use a modified blackbody: 
\begin{equation}
I_{\nu} \propto Q_{0}\left(\frac{\nu}{\nu_{0}}\right)^{\beta}B_{\nu}(T_{\rm{dust}})
\end{equation}
where $\nu_{0}$ is the frequency at which the temperature is calculated, where $Q_{0}/\nu_{0}^{\beta}$ is a constant, and where $B_{\nu}$ is the Planck function. We take the power-law index $\beta$ equal to 2, as in \citet{lau2013} and consistent with extinction curves from \citet{draine2003}. We separately do the same calculation assuming $\beta$ = 0 (blackbody). The temperature is therefore calculated following the equation:
\begin{equation}
\rm{L}^{\prime} - \rm{Ms} = -2.5\textrm{ log}\left[\left(\frac{\nu_{\rm{L}^{\prime}}}{\nu_{\rm{Ms}}}\right)^{\beta}\frac{B_{\rm{L}^{\prime}}(T_{\rm{dust}})}{B_{\rm{Ms}}(T_{\rm{dust}})}\right]
\end{equation}

From our $\rm{L}^{\prime}$ and Ms measurements, we are able to obtain a dereddened color ($\rm{L}^{\prime}$ - Ms) of 0.706. Fitting a modified blackbody following equation 2 with $\beta$ = 2, the color temperature we obtain from our 2005 data is equal to 568 $\pm$ 44 K; assuming a blackbody ($\beta$ = 0), we obtain a 2005 temperature of 426 $\pm$ 44 K where our error bars are computed via a Monte Carlo simulation.

Using our $\rm{L}^{\prime}$ and Ms photometric data in 2016 when G1 is observed to be point-like and assuming that G1 behaves as a blackbody in this epoch ($\beta$=0), we infer a blackbody temperature of 684 $\pm$ 75 K (where our error bars are again computed via a Monte Carlo simulation). Therefore, our inferred blackbody temperature has increased from 2005 to 2016.

\section{Discussion}\label{discussion}
G1 is a cold, extended source that has tidally interacted with Sgr A$^{*}$ and survived at least 13 years past periapse passage. It has observable parameters that seem to be consistent with other examples of infrared excess sources at the Galactic Center, the most prominent of which is G2. Many of its orbital and observable properties are comparable with those of G2: its cold temperature (426 K if $\beta$ = 0, or 568 K if $\beta$ = 2 in 2005; 684 $\pm$ 75 K if $\beta$ = 0 in 2016), its highly eccentric orbit ($e$ = 0.99$^{+0.001}_{-0.01}$), and the orientation of the orbital plane (see Table~\ref{tab:orbital_solution}). There is a measurable size change post-periapse passage, and the $\rm{L}^{\prime}$ flux density also changes dramatically after periapse. In the following, we discuss the similarities and differences between G1 and G2.

\subsection{Is G1 part of a gas streamer common with G2?}\label{gasstr}
\citet{pfuhl2015} have recently proposed that G1 and G2 are not only lying in the same orbital plane, but follow the same trajectory. They speculate that the Keplerian orbits of G1 and G2 are closely related and they postulate the small deviations between the orbits of the two objects are due to the drag force from the ambient Galactic Center medium. This additional drag force leads to an evolution of G2's orbit into G1's orbit over time. Similarly, \citet{mccourt2015} and \citet{madigan2016} use G1 and G2 as probes to constrain the properties of the accretion flow surrounding Sgr A$^{*}$. They model the orbital differences (as found by \citealt{pfuhl2015}) between G1 and G2 in terms of an interaction with the background flow \citep{mccourt2015} and in the accretion flow onto Sgr A$^{*}$ \citep{madigan2016}. Based on their orbital analysis, they conclude that both sources could have originated from the clockwise young stellar disk \citep{paumard2009, lu2009, yelda2014}. 

However, the study we present here, which includes data taken several years beyond the last data point used in \citet{pfuhl2015} (2014.6 vs. 2010.5; true anomalies of 10.5 and 8.7 degrees, respectively), shows that despite the common orbital plane, G1 and G2 have distinct Keplerian orbits with a significant ($>$3$\sigma$) difference of their arguments of periapse, $\sim$3 times larger than the difference reported in \citet{pfuhl2015}. This is demonstrated in Fig.~\ref{fig:comp_orbit}, showing both the data and the best-fit orbits projected into the plane of the sky as well as both best-fit orbits projected into the average orbital plane.

{Our findings do not firmly exclude the models proposed by \citet{mccourt2015} and \citet{madigan2016}. However, while both models might be able to accommodate such a large change of the Keplerian orbit in the case of a compact gas cloud, the drag force scenario and a resulting common trajectory of G1 and G2 become increasingly unlikely in the context of a central star and thus larger object masses, as indicated by the compactness and brightness of both sources.} The masses derived in the following sections and in \citealt{witzel2014} are $10^{5}$ - $10^{6}$ times larger than the originally proposed 3 Earth masses. The interpretation in \citet{pfuhl2015} that G1 and G2 are two dense regions within the same extended gas streamer that fills one trajectory around the black hole and have an identical origin, but are offset by $\sim$13 years, therefore seems unlikely.

The orbital planes of G1 and G2 are very similar and they are fairly close to the plane defined by the clockwise disk (\citealt{yelda2014} and see Fig.~\ref{fig:2d_posterior} of \citealt{pfuhl2015}). G1 and G2 may have therefore originated the clockwise disk. We note, however, that there are other G2-like sources that do not lie on their common orbital plane \citep{Sitarski:2015aa}.

\subsection{Evolution of G1's Dust Envelope}\label{evolG1}

Independently of whether G1 and G2 are related by a gas streamer, the physical natures of G1 and G2 are still not yet known. Recent results (e.g., \citealt{witzel2014}, \citealt{valencia2015}, in contrast to \citealt{pfuhl2015}) support the hypothesis that that G2 has a stellar component due to its periapse passage survival. This raises the question of whether there is similar evidence that G1 is stellar in nature.

In contrast to observation that G2 is unresolved at $\rm{L}^{\prime}$, for G1 we are able to measure its size in 2005 and we can therefore put constraints on the optical depth, $\tau$, of the dust envelope at this point in time. Based on several parameters calculated in Section~\ref{phottemp} ($T_{\beta = 2}$ = 568 K; $T_{\beta = 0}$ = 426 K; $r_{\rm{G1}, 2005}$ = 137 AU), we find that the optical depth of G1 is small in the epochs when it is resolved, and we can therefore conclude that the origin of the extended continuum emission is an optically thin medium in 2005. As calculated in Section~\ref{specanal}, the ambient radiation field in the Galactic Center is strong enough (with Lyman-$\alpha$ alone) to externally heat this optically thin shell. The profile of G1 in the epochs where it is extended is well constrained by a PSF convolved with a 2D Gaussian (see Section~\ref{photsize}) and shows no evidence of two components (as could be modeled by a PSF + a 2D Gaussian). This indicates that we do not see a central, optically-thick point source in 2005.

\begin{figure}
\begin{center}
\includegraphics[width=8cm]{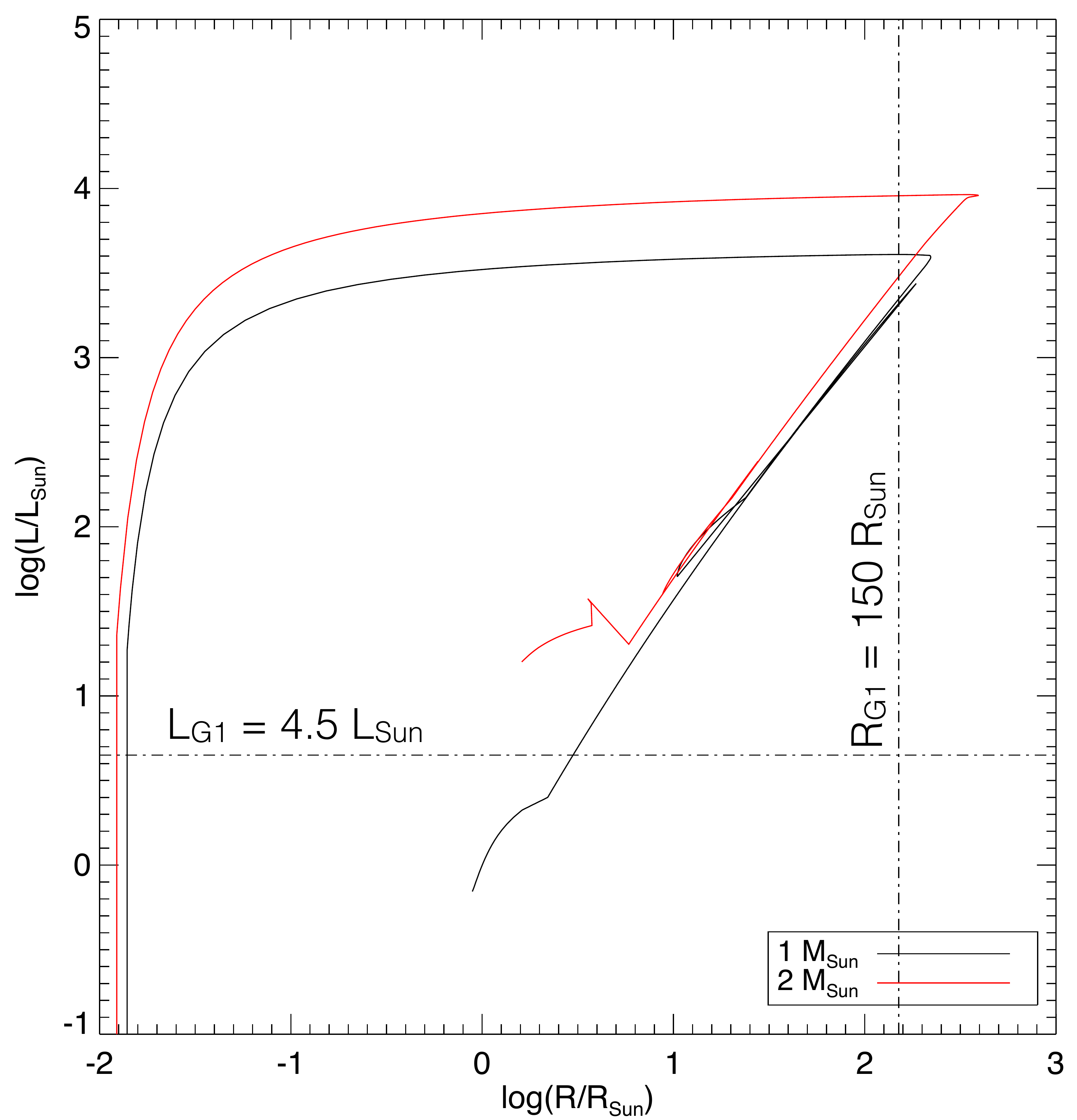}
\caption{Evolutionary tracks of stars of 1 (black) and 2 (red) $\Msun$ in the luminosity-radius plane, computed using the SSE code \citep{hurley2000}. The vertical and horizontal lines show the inferred blackbody values for G1 from our 2016 data set. It is clear that the luminosity we infer for G1 is too small for a source we see of that radius.}\label{fig:gunther_figure_3}
\end{center}
\end{figure}

From 2009 onwards, G1 is unresolved at $\rm{L}^{\prime}$ and shows a significantly lower, roughly constant flux density of $\sim$0.6 mJy. Blackbody modeling of G1's $\rm{L}^{\prime}$ - Ms color yields a temperature of 684 K, implying a blackbody radius of $\sim$1 AU and a luminosity of $\sim$4.5 solar luminosities. This high luminosity and the fact that the object become more compact with time point to a substantially larger mass than 3 Earth masses. As indicated by the evolutionary tracks of main sequence stars, this mass can be of the order of a solar mass (Fig.~\ref{fig:gunther_figure_3}). However, the large derived blackbody size for the unresolved G1 show that it is not a main sequence star nor is G1 luminous enough to be a Red Giant.

\begin{figure}
\begin{center}
\includegraphics[width=8.0cm]{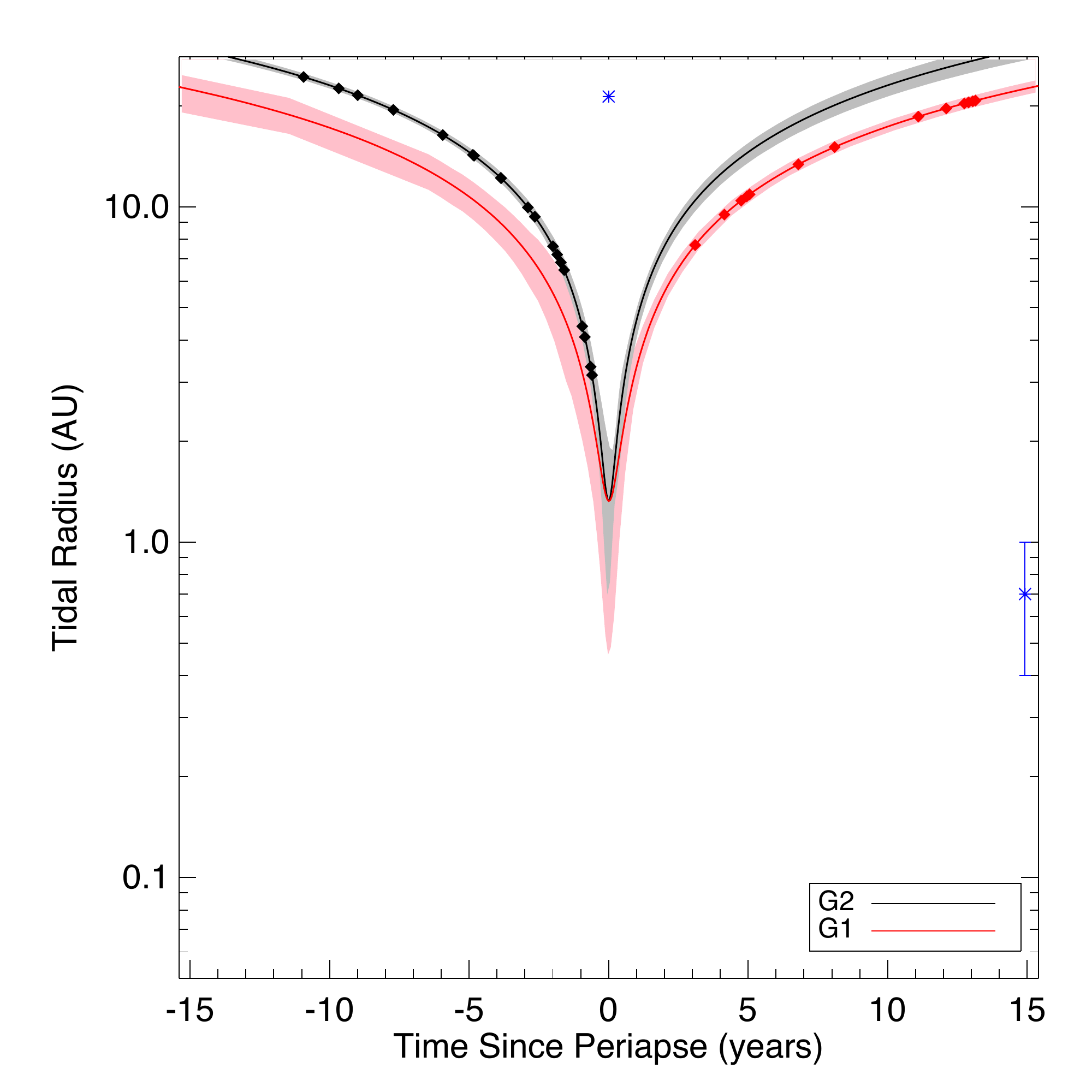}
\caption{Plot of the tidal radii of G1 (red) and G2 (black). The tidal radius for G1 is computed assuming a mass of 1.0 $\Msun$, consistent with our luminosity calculation described in Section~\ref{evolG1}. The first blue asterisk denotes the inferred size of G1 at periapse passage calculated with our dynamical model; the second blue asterisk shows the inferred size assuming that G1 is a blackbody in 2016. The latter is consistent with 1.0 AU, the deepest point of direct tidal interaction of G1 with Sgr A$^{*}$. G1 has a longer interaction with Sgr A$^{*}$ than G2 does.}\label{fig:tidal}
\end{center}
\end{figure}

\begin{figure*}
\begin{center}
\includegraphics[width=11.0cm]{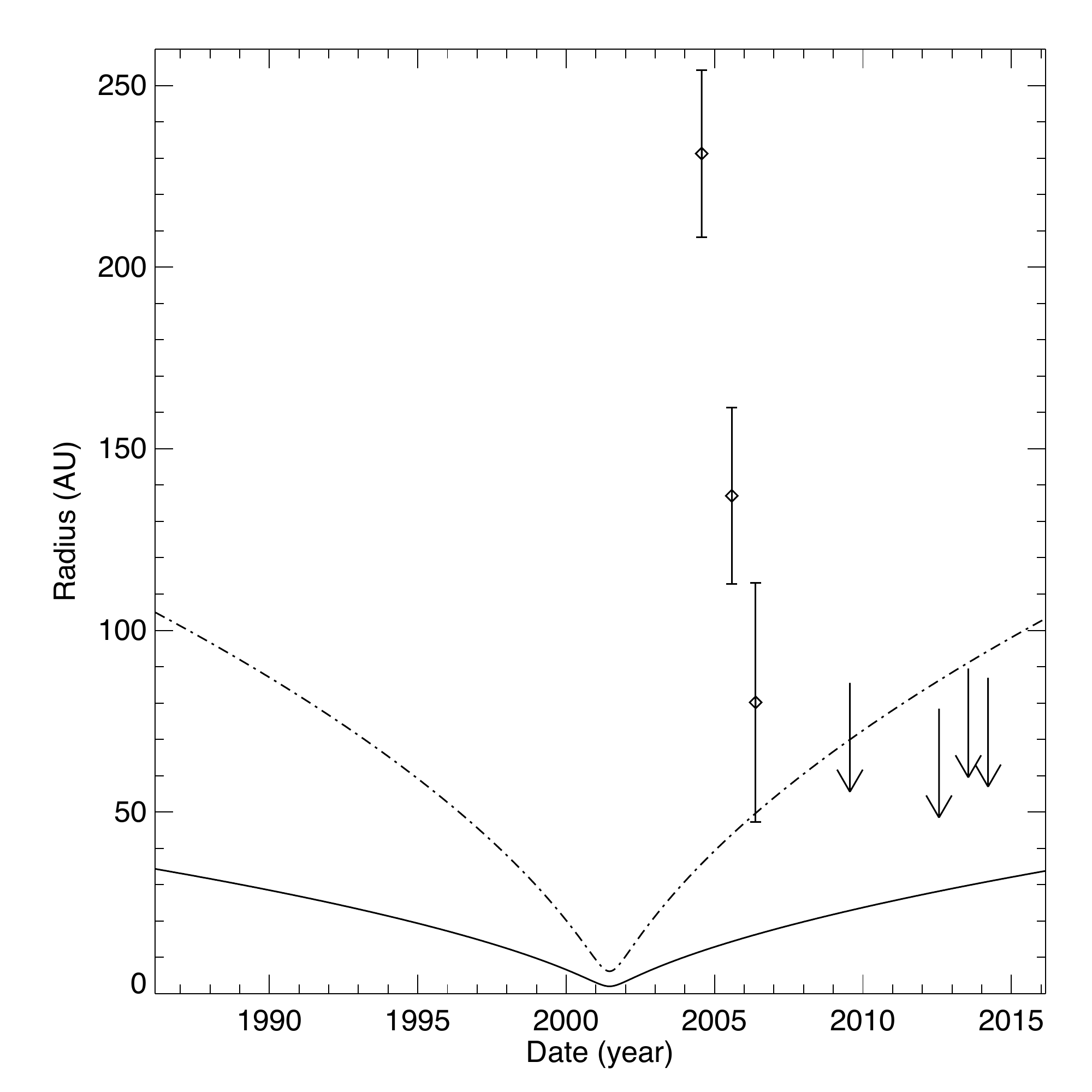}
\caption{Tidal radius as a function of time since periapse passage for G1. The solid line shows the tidal radius of a 2$\Msun$ main-sequence star (as found for G2 in \citealt{witzel2014}); the dashed line shows the tidal radius of a 100$\Msun$ star. The intrinsic size of the semi-major axes of G1 from Fig.~\ref{fig:size} are over-plotted as well. It is evident that in the epochs where G1 is large, it lies well outside the tidal radius and therefore can interact gravitationally with Sgr A$^{*}$. Therefore, some of the dust evolution could have become unbound, but the remainder survives as a compact object in later epochs (and when the tidal radius is outside the size of the source).}\label{fig:tidrad}
\end{center}
\end{figure*}

The material at the enormous radial distance from the center of G1 of $r \sim$ 230 AU of the outer halo seen in the extended epochs certainly remains unbound from G1 for even much higher G1 masses than 1 $\Msun$; in fact, this holds true for a central mass that is two orders of magnitude higher due to the weak $M^{1/3}$ dependence of the tidal radius (Fig.~\ref{fig:tidal} shows the tidal radius (black lines) of a 2$\Msun$ source (solid line) and a 100$\Msun$ source (dashed line; see \citealt{witzel2014}) plotted with the measured HWHMs of G1). Therefore this material is stripped and its emission falls below the detection limit as its density decreases or grains are destroyed by X-rays and high-energy particles generated in the accretion flow (e.g., \citealt{lau2015} and references therein; \citealt{tielens1994}). It is interesting to note that the minimal radius of material that remained bound throughout periapse for a G1 mass of 1 $\Msun$ and the periapse passage distance of $\sim$300 AU is 1 AU (see Fig.~\ref{fig:tidrad}, which plots the tidal radius as a function of time since periapse passage for G1 and G2). This corresponds nicely to the derived blackbody radius in 2016.

The question remains how G1 has reached the enormous extent of $d$ = 460 AU in 2004. This likely requires that G1 was already large at periapse passage. From an energy argument, we can determine the size of G1 at periapse passage from the maximum shearing velocity, $v_{\rm{sh}}$ of the object in the potential of the black hole according to the following equation:
\begin{equation}
\begin{split}
v_{\rm{sh}}^{2} = \left(\frac{r_{\rm{obs}} - r_{\rm{per}}}{\Delta t}\right)^{2} = \\
\left[\sqrt{v_{*}^{2} - 2GM_{\rm{BH}}\left[\frac{1}{d_{*}} - \frac{1}{d_{*}-r_{\rm{per}}}\right]} - v_{*}\right]^{2} - \frac{2Gm}{r_{per}}
\end{split}
\end{equation}
where $r_{\rm{obs}}$ is the observed size in 2004, $r_{\rm{per}}$ is the half-width along the Sgr A$^{*}$-G1 line, $v_{*}$ is velocity of G1 at periapse passage, $m$ is the mass of G1, $M_{\rm{BH}}$ is the mass of the supermassive black hole from Table~\ref{tab:orbital_solution}, $\Delta t$ is the difference between our observation date (2004.6) and periapse passage time, the first epoch where we see a resolved G1, and $d_{*}$ is the distance of the center of G1 to the Sgr A$^{*}$. Simultaneously solving for $m$ and $r_{\rm{per}}$, we find that $r_{\rm{per}}$ is larger than 21 AU at the time of closest approach and that the solution is not strongly mass dependent.

Therefore, for G1 to have its measured size be so large in 2004, $r_{\rm{per}}$ must be $\ge$21.3 AU at periapse passage and G1 was likely a large object even before it started interacting with the SMBH. If G1 began with a radius of 4, 3, or 2 AU, it would have started interacting tidally with the SMBH 1.3, 0.9, and 0.4 years before periapse passage, respectively, giving it plenty of time to grow to be the large source we infer for periapse passage.

While it is possible that G1 appears extended at $\rm{L}^{\prime}$ because of confusion with background sources, we judge this to be unlikely. We have traced the orbits of all known stars close to G1 and Sgr A$^{*}$, and G1 is certainly not confused with a bright (mag$_{\rm{L}^{\prime}}$ $<$ 16) source. But it is not fully excluded that, during the early epochs, there could be several dim stars whose images are overlapping that of nearby G1 for multiple epochs before separating and moving below the detection limit again. However, the symmetry in the extended residual after subtracting a point source makes this seem rather unlikely.

\subsection{The 'bloated star' model}

Our suggested model for G1's dust shell is as follows: G1 started tidally interacting with the SMBH with a rather large size several years prior to periapse passage. The tidal radius penetrated deep into the dust shell ($r \sim$ 1 AU) and the outer part of the optically thick shell became unbound from the source. This unbound shell became optically thin and externally heated by the surrounding radiation field in the Galactic Center, which is what we observe starting in 2004. Over time, the tidally stripped dust expanded away and fell below the detection limit against the local background emission, and by 2009, we see the optically thick surface of a massive, internally-heated object as a point source that is 2 magnitudes fainter than what is observed in 2004. Throughout all epochs, the source is also surrounded by an externally-heated gas envelope that we observe as Br-$\gamma$ emission. As we discuss below in Section~\ref{compmerge}, a possible physical explanation for G1's large size is that it could be an example of a black-hole-driven binary merger product \citep{phifer2013, witzel2014, prodan2015}.

\subsection{Comparison to G2: Gas Cloud or Star?}

Several predictions have been made for the post-periapse development of G2 in the case of a pure gas cloud. G1 and G2 have similar periapse passage distances and blackbody sizes (as inferred in Section~\ref{phottemp} for G1), and we expect them to tidally interact with the black hole in a comparable manner. Thus, in the following, we compare G1's post-periapse observables to some of these predictions for G2.

Various models for G2 predict that if it were a pure gas cloud, it should undergo tidal shearing within 1 to 7 years after periapse. The Br-$\gamma$ flux of G2 was predicted to rapidly decrease over time \citep{anninos2012, morsony2015}, both due to the break-up of G2 and the heating of its gas. Observationally, the latest Br-$\gamma$ line detection of G1 occurred in 2008, 7 years after periapse passage \citep{pfuhl2015}\footnote{In epochs later than 2008, the Br-$\gamma$ line is extremely difficult to extract due to lack of sufficient data quality.}, not showing any indication of a strong decay or complete depletion. In fact, the post-periapse luminosity of G1 is consistent with the pre-periapse luminosity of G2 (see Section~\ref{specanal}). We also note that G1's FWHM in 2006, 5 years after periapse passage, was 185 km sec$^{-1}$, comparable to the line width of G2 five years before its periapse passage in 2014. \citep{phifer2013} This provides strong constraints on future hydrodynamic modeling of the post-periapse development of these objects.

Unlike G2's flux density staying constant before and during periapse \citep{witzel2014}, G1's $\rm{L}^{\prime}$ flux significantly decreased post-periapse (Fig.~\ref{fig:phot}). The size of G1 at $\rm{L}^{\prime}$ shows a similar development over time from a clearly resolved, optically thin source two years after periapse to an unresolved, compact source five years post-periapse passage. Our calculation in the previous section indicates that G1 went through periapse passage with a radius  $> 21$ AU. These findings are indicative of G1's dust envelope interacting more strongly with Sgr A$^{*}$ than that of G2 due to its smaller mass and larger size. While they have similar tidal radii close to periapse passage, G1 interacts with the SMBH for a longer period of time than G2.

Several studies (e.g., \citealt{schartmann2012, anninos2012, gillessen2012, gillessen2013a, morsony2015}) find that if G1 or G2 were a gas cloud, there should be a significant increase in the steady-state X-ray flux several months before and after periapse passage due to shocks. The \textit{Chandra X-ray Observatory} was launched in 1999, and the earliest observations of Sgr A* were conducted in late 1999 and 2000. \citet{baganoff2001} and \citet{ponti2015} show no indication of an increase in the the steady-state X-ray flux in the time around G1's periapse passage (2001.3 $\pm$ 0.4; Fig. 3 in \citealt{ponti2015}), although they raise the possibility that there was an increase in the rate of bright flares.

The size evolution of G1 in $\rm{L}^{\prime}$, along with the distinct Keplerian orbits described in Section~\ref{gasstr}, the intact Br-$\gamma$ emission after 7 years, the survival of G1 at $\rm{L}^{\prime}$, and the lack of an increased X-ray flux, all provide evidence that G1 has a massive ($\sim$1 \Msun) central (stellar) component surrounded by an envelope of gas and dust, similar to our hypothesis for G2 \citep{witzel2014}.  Even if the mass of G1 is smaller than 1 $\Msun$, it is still $\sim10^{5}$ times larger than the masses suggested for a gas cloud \citep{gillessen2012, pfuhl2015}.

\subsection{Comparison to Observed Merged Binary Systems}\label{compmerge}
G1 shares some observed characteristics with known merged binaries. Our inferred dust temperature of 426-568 K is within the ranges reported for other observed binary mergers, including V1309 Sco \citep{nicholls2013} and BLG-360 \citep{tylenda2013}. Also, the large size and luminosity inferred for G1 are similar to BLG-360. We do suggest that G1 and G2, if they are indeed binary mergers, crossed their individual Roche limits sometime between 1 $\times$ 10$^{4}$ - 1 $\times$ 10$^{6}$ years after the last star formation episode \citep{stephan2015}. The high eccentricity of G1 and G2 in their respective orbits around Sgr A$^{*}$ is what we would expect from binary systems that have been affected by the Kozai mechanism \citep{kozai1962, lidov1962}. That is, we do not assume that these mergers stem from random stellar collisions, but rather that the binaries merge as a result of secular interactions similar to what is described by \citet{prodan2015} and \cite{stephan2015}. The end result of the eccentric Kozai mechanism yielding a merger product has been discussed in detail in the literature (see the review by \citealt{noaz2016}); binary systems are most likely to merge on highly eccentric orbits \citep{naoz2014, stephan2015}.
%We do suggest that G1 and G2, if they are indeed binary mergers, merged sufficiently long ago for the merged system to have dynamically settled into a quasi static configuration. 

Merged binary systems undergo many physical changes as the merger occurs. For example, there is usually an optical outburst immediately following the physical merging of the stars, along with an evolution of spectral type (e.g., \citealt{tylenda2011}, \citealt{nicholls2013}). The very few examples that have been published thus far have been inferred to be merged binaries because of optical periodic variability from the binary system before the outburst, and the absence of any periodicity from the system following the outburst (e.g., \citealt{tylenda2011}). 

The high infrared flux density of G1 (2.7 mJy at $\rm{L}^{\prime}$) shortly after periapse passage (2005) is consistent with the high fluxes from other binary merger products after the merger has taken place. As the majority of stars in the field and in dense stellar clusters like the nuclear star cluster exist as multiple-component systems (e.g., \citealt{prodan2015, sana2011,duchene2013}), it is not unreasonable that many of these could merge in the Galactic Center and form extended envelopes of gas and dust.

While the binary merger hypothesis provides many similarities to the observed characteristics of G1, several things remain unclear. The timescale over which such mergers occur is not yet known (but is under study by \citealt{stephan2015}); the length of the dusty phase depends on the mass of the progenitors and the relaxation timescale. For instance, V1309 Sco was originally discovered in September 2008 as a ``red nova" \citep{nakano2008, rudy2008a, rudy2008b,tylenda2011} that had an evolving spectral type from F to M. \citet{nicholls2013} showed that V1309 Sco was undetected in the near-infrared regime prior to its outburst;  $\sim$23 months afterward, there was a clear near- and mid-infrared excess. They further model the infrared excess as a dust envelope surrounding V1309 Sco that formed after the merging. Two years after merging, a near-infrared excess was still present \citep{nicholls2013}. This implies that the duration of the dusty phase for G1 was at least 13 years so that our observation window is shorter than the duration of the dust phase. However, if G1 and G2 are more massive sources, winds and radiation stemming from the star could affect the dust envelope lifetime. Several other hypotheses exist that could describe the observables of G1, such as edge-on, protoplanetary disks around young, low-mass stars \citep{murrayclay2012}, disks around older stars \citep{miralda2012}, or some other tidal disruption phenomenon involving a stellar object.

{Nonetheless, the probability of finding two merger products on similar orbits is not negligible. The argument is twofold: The first considers the probability to find mergers that take place now, and the other the orbital parameters of such mergers:} {(1) \cite{stephan2015} calculated the percentage of expected mergers to be rather high at about 10\% of all binary star systems within the first few million years after the latest star formation event.
(2) While the two orbits have similar eccentricities, \citet{stephan2015} has shown that there is reason to believe that it is more probable to encounter merger products at high eccentricities which is consistent with the underlying physics of the eccentric Kozai-Lidov mechanism (\citealt{noaz2016}). It is additionally likely to discover G-like objects on high eccentricities within our observational constraints, given that they closely approach the black hole and are fast-moving objects. Furthermore, the similarity of the orbital orientations could be explained with a common origin, for example in the clockwise disk. In summary, the binary merger scenario appears to be a plausible interpretation of our observations.}

\section{Conclusions}\label{concl}
G1 has several observable properties similar to those of the mysterious G2 object--it is a cold source in the Galactic Center that has hydrogen recombination emission (at Br-$\gamma$) and has recently passed very close to Sgr A$^{*}$. Our orbital fits indicate that G1 and G2 lie on similar orbital planes, but have different arguments of periapse, indicating that these objects are not part of the same gas streamer. In contrast to G2, G1 was originally well-resolved at $\rm{L}^{\prime}$ (3.8 $\mu$m). This additional information strongly supports the idea that there is a central, stellar object embedded in a gas- and dust-filled envelope. 

We hypothesize that G1 may be a binary merger product due to the similarities to observed merger systems (see Section~\ref{evolG1}): notably, it has a large inferred size, and high infrared luminosity. This would be a natural explanation for many unsolved questions regarding other populations in the Galactic Center, including the young stars in the S-star cluster, which may have resulted from the mergers of binaries interacting with Sgr A$^{*}$, followed by relaxation back to the main sequence. G1 and G2 are also not the only objects with these observed properties in the Galactic Center, as at least 4 others exist close in proximity to Sgr A$^{*}$ \citep{sitarski2014}. Further studies of these additional sources will indicate whether all these sources have common characteristics such as Br-$\gamma$ emission, and whether they share a common origin or a common production mechanism.

\acknowledgements{Support for this work was provided by NSF grants AST-0909218 and AST-1412615, the Levine-Leichtman Family Foundation, the Preston Family Graduate Fellowship (held by B. N. S. and A. B.), and the UCLA Graduate Division Dissertation Year Fellowship (held by B. N. S.). The W. M. Keck Observatory is operated as a scientific partnership among the California Institute of Technology, the University of California, and the National Aeronautics and Space Administration. The authors wish to recognize that the summit of Mauna Kea has always held a very significant cultural role for the indigenous Hawaiian community. We are most fortunate to have the opportunity to observe from this mountain. The Observatory was made possible by the generous financial support of the W. M. Keck Foundation. B. N. S. also thanks James Larkin for his engaging discussions about the nature of G1 and G2, Alexander Stephan for his comments, and Ann-Marie Madigan for her thoughts on G1. We thank Arezu Dehghanfar for help with preparing the manuscript. We thank the anonymous referee for providing invaluable comments.}

\appendix
\section{$\rm{K}^{\prime}$-derived distortion solution on $\rm{L}^{\prime}$ data}\label{app1}
\begin{figure}
\begin{center}
\includegraphics[width=16cm]{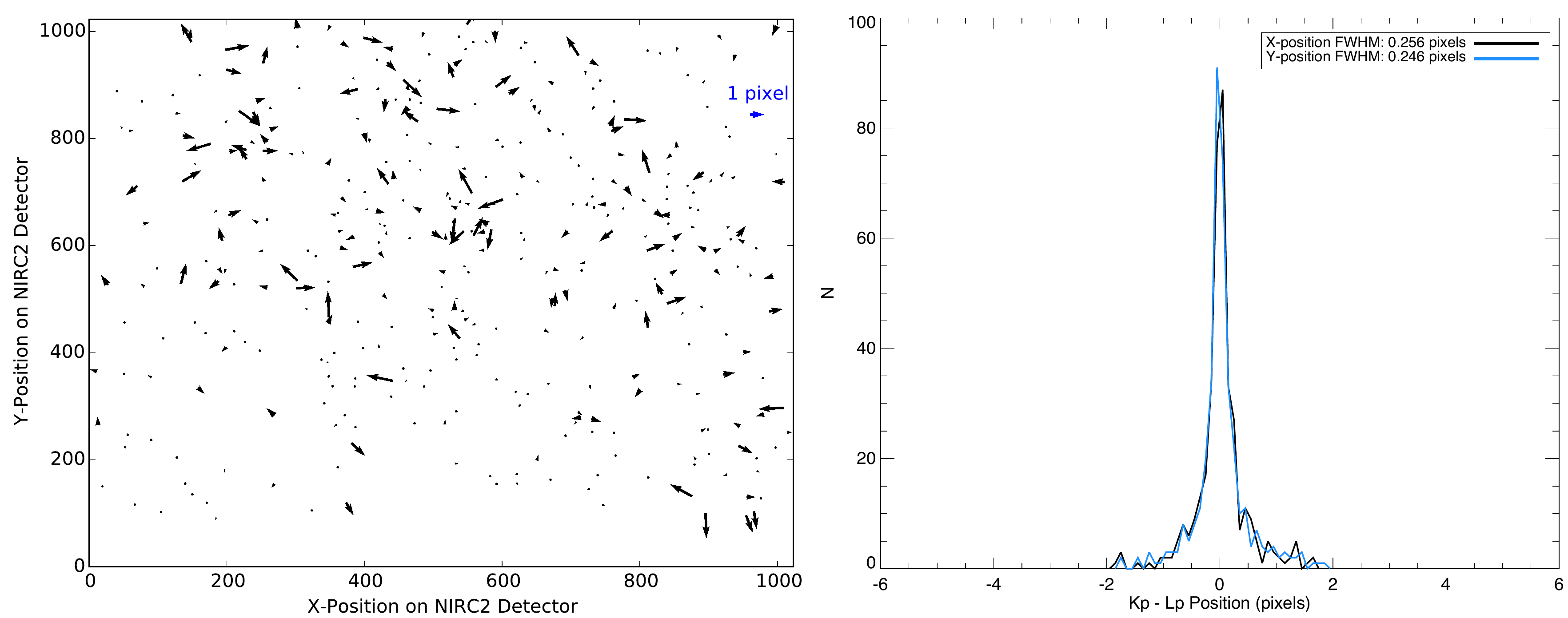}
\caption{\textit{Left}: Difference between the $\rm{K}^{\prime}$ and $\rm{L}^{\prime}$-transformed positions across the field of view of NIRC2. The arrows originate at the $\rm{K}^{\prime}$ position and end at the position of the $\rm{L}^{\prime}$-transformed-to-$\rm{K}^{\prime}$ coordinate system points. There is no preferential position or rotation of the arrows, so using the $\rm{K}^{\prime}$ data derived distortion solution is completely adequate. \textit{Right}: Histogram of the difference between the $\rm{K}^{\prime}$ and $\rm{L}^{\prime}$-transformed positions. The FWHMs of the distributions are less than the astrometric errors of the data.}
\label{fig:distortion}
\end{center}
\end{figure}

As stated in Section~\ref{datasets}, we resample all data ($\rm{K}^{\prime}$, $\rm{L}^{\prime}$, and $\rm{Ms}$) with the geometric optical distortion solution from \citet{yelda2010}. This distortion solution was derived with $\rm{K}^{\prime}$ data only, so we tested whether this distortion solution was inappropriately applied to the $\rm{L}^{\prime}$ and $\rm{Ms}$ data sets. We therefore took one of our epochs of data (we chose 2005.580) and transformed the $\rm{L}^{\prime}$ positions as detected by \textit{StarFinder} (see Section~\ref{astrometry}) into the $\rm{K}^{\prime}$ coordinate system. We allowed for first-order translation, rotation, and pixel scale adjustments between the two frames that were independent in $x$ and $y$. The results from this alignment are shown in the left panel of Fig.~\ref{fig:distortion} where each arrow represents the difference in position for stars identified both at $\rm{K}^{\prime}$ and $\rm{L}^{\prime}$. As there is no noticeable rotation or structure indicated by arrows, we conclude that applying the distortion correction for $\rm{L}^{\prime}$ data is therefore adequate. The right panel of Fig.~\ref{fig:distortion} shows a histogram of the difference between the $\rm{K}^{\prime}$ and $\rm{L}^{\prime}$-transformed coordinates in both $x$ and $y$. The FWHMs of these histograms are less than the positional errors as found in Tab.~\ref{dataG1}.

\section{Jackknife-derived estimates of the variance of G1's orbital parameters and the black hole parameters}\label{jackg1}
To determine whether our errors for the orbit of G1 and the black hole parameters capture at least part of the systematic errors due to potential outliers, we used a jackknife resampling technique to determine the variance of each of G1's orbital parameters while simultaneously fitting S0-2, S0-38, and G1. In each of our orbital fits, we dropped one epoch of observations and determined the jackknife variance over all orbital fits. Our recovered jackknife parameters are listed in Tab.~\ref{jackn}. The values and associated error bars calculated from this jackknife analysis are consistent with what is reported in Tab.~\ref{fitBHpar}.

\begin{deluxetable}{ll}
\tabletypesize{\scriptsize}
\tablecaption{Jackknife Parameters \label{jackn}}
\tablewidth{0pt}
\tablehead{\colhead{Orbital Parameter} & \colhead{Jackknife Parameters}}
\startdata
X-Position of Sgr A* ($x_{0}$, mas) & 2.2$\pm$0.3\\
Y-Position of Sgr A* ($y_{0}$, mas) & -4.3$\pm$0.3\\
$\Delta$RA Velocity of Sgr A* ($V_{x}$, mas/yr) & 0.11$\pm$0.02\\
$\Delta$Dec Velocity of Sgr A* ($V_{y}$, mas/yr) & 0.67$\pm$0.03\\
Radial Velocity of Sgr A*($V_{z}$, km/sec) & -19.3$\pm$3.7\\
Distance to Sgr A* ($R_{0}$, kpc) & 7.85$\pm$0.06\\
Mass of Sgr A* ($M$, millions of $\Msun$) & 3.92$\pm$0.06\\
\hline
\hline
G1 Parameters:\\
\hline
\hline
Periapse Passage Distance($a_{\textrm{min}}$, AU) & 292$\pm$44\\
Time of closest approach ($T_{0}$, years) & 2001.3 $\pm$0.2 \\
Eccentricity ($e$) & 0.993$\pm$0.002\\
Argument of periapse ($\omega$, degrees) & 117$\pm$4\\
Inclination ($i$, degrees) & 109$\pm$1\\
Position angle of the ascending node ($\Omega$, degrees) & 86$\pm$6
\enddata
\end{deluxetable}

\bibliographystyle{apj}
\bibliography{g1_paper}

\end{document}